\documentclass[conference,compsoc]{IEEEtran}
\usepackage{xcolor}
\usepackage{tabularx}
\usepackage{booktabs}
\usepackage[nocompress]{cite}
\usepackage{stfloats} 
\usepackage{multirow}
\usepackage{url}
\usepackage[group-separator={,}]{siunitx}
\usepackage[shortlabels]{enumitem}

\ifCLASSINFOpdf
   \usepackage[pdftex]{graphicx}
\else
\fi

\ifCLASSOPTIONcompsoc
 \usepackage[caption=true,font=footnotesize,labelfont=sf,textfont=sf]{subfig}
 \usepackage{float}
\else
 \usepackage[caption=true,font=footnotesize]{subfig}
 \usepackage{float}
\fi

\begin{document}

\title{On Generalisability of Machine Learning-based \\Network Intrusion Detection Systems}

\author{\IEEEauthorblockN{Siamak Layeghy\IEEEauthorrefmark{1}, 
Marius Portmann\IEEEauthorrefmark{2}}
\IEEEauthorblockA{School of ITEE\\
University of Queensland\\
Brisbane, QLD 4072, Australia\\
\IEEEauthorrefmark{1}siamak.layeghy@uq.net.au, 
\IEEEauthorrefmark{3}marius@ieee.org
}
}

\maketitle

\begin{abstract}
%
%
Many of the proposed machine learning (ML) based network intrusion detection systems (NIDSs)  achieve near perfect detection performance when evaluated on synthetic benchmark datasets. 
%
Though, there is no record of if and how these results generalise to other network scenarios, in particular to real-world networks.
In this paper, we investigate the generalisability property of ML-based NIDSs by extensively evaluating seven supervised and unsupervised learning models on four recently published benchmark NIDS datasets. 
Our investigation indicates that none of the considered models is able to generalise over all studied datasets.
Interestingly, our results also indicate that the generalisability has a high degree of asymmetry, i.e., 
%
swapping the source and target domains can significantly change the classification performance.
Our investigation also indicates that overall, unsupervised learning methods generalise better than supervised learning models in our considered scenarios.
Using SHAP values to explain these results indicates that the lack of generalisability is mainly due to the presence of strong correspondence between the values of one or more features and Attack/Benign classes in one dataset-model combination and its absence in other datasets that have different feature distributions.  

%

\end{abstract}
\IEEEpeerreviewmaketitle

\section{Introduction}
A quick search of the academic literature in network intrusion detection systems (NIDSs) reveals that there are hundreds of proposals based on machine learning (ML) algorithms.
In most of these proposals, the performance is evaluated using a publicly available benchmark NIDS dataset such as UNSW~\cite{unsw}, CIC-IDS~\cite{cic}, and KDD99~\cite{KDD99}. 
For this purpose, like in other fields that use machine learning, the benchmark dataset is divided into the training and test subsets, the ML models are trained on the training subset and evaluated on the test subset.
In some of these studies, such as ~\cite{sarhan2020netflow} and \cite{egraphsage}, multiple benchmark dataset are used in this way. That is, the ML model is trained and tested on each benchmark dataset separately. 

While the accuracy, detection rate and other performance metrics reported in many of these studies are near perfect, and the studies are designed and performed rigorously, these excellent results have unfortunately not translated into  ML-based NIDSs in practical network deployments with close to 100\% detection performance. 

We believe that an investigation into the generalisability properties of ML-based NIDSs is an important, but so far under-investigated step towards bridging the gap between the excellent results achieved by the academic research community, and the practical impact of the research.    
%
%
%
In particular, we assume that the setup/environment in which benchmark NIDS datasets are created/collected varies among the different available benchmark datasets, and most likely also varies significantly from the characteristics of potential real-world deployment networks.

Applying machine learning for detecting unseen datasets and unseen classes of the same datasets has been explored in many previous works. This includes studies that investigate zero-day/unseen attack detection such as~\cite{sarhan2021zero} and \cite{Rivero2017} or studies that examine domain adaptation and transfer learning in the field of ML-based NIDSs such as~\cite{Zhao2019} and \cite{Wu2019}.
Though, none of these works specifically investigates the generalisability in ML-based NIDSs.
Perhaps the only previous works that partially/indirectly  investigate the generalisability of ML-based NIDSs are ~\cite{Al-riyami2018} and \cite{Apruzzese2022} in which the cross-evaluation of ML-based NIDSs has been explored.
Apart from these studies that have been discussed in the related works of this paper, we have not been able to find other works considering the generalisability of these NIDSs.

Our work aims to address this gap by taking advantage of four recently published NIDS benchmark datasets~\cite{sarhan2021towards}, which have been converted from their original format into a common NetFlow-based format, with an identical feature set.
We evaluate the generalisability in both directions across different datasets, i.e., each dataset is used as the training and test dataset, against all other datasets. 
In addition, we evaluate the generalisability of both supervised as well as unsupervised learning models. 

Then we compare the results of single domain evaluation, in which the same benchmark dataset is used for both training and evaluation, with the results of generalisability evaluation in which a trained model is evaluated against the  other three benchmark datasets.
Finally, we apply the Shapley values~\cite{shapley1953value} on data and models in different scenarios and explain the generalisability by comparing the Shapley values for different experiments.

\section{Related Works}\label{related works}

While there is no previous work systematically evaluating the generalisability of ML-based NIDSs, there are two previous studies that partially investigate evaluation of ML-based NIDSs via datasets unseen during training which include~\cite{Al-riyami2018} and \cite{Apruzzese2022}.

In~\cite{Al-riyami2018} it is shown that the near perfect performance of simple supervised ML models on the dataset used for their training considerably drops when evaluated on other, unseen benchmark datasets.
This study uses the Kyoto+~\cite{Kyoto}, gureKDD~\cite{gureKDD} and NSL-KDD~\cite{NSL-KDD} NIDS datasets which are published in 2006, 2008 and 2009 respectively. 
However, the Kyoto+ dataset is not included in the evaluation of unseen datasets, due to the difference in the feature set with other datasets.
While this study discusses cross evaluation of ML-based NIDSs, it does not investigate the factors for the poor performance on the ML-models on unseen benchmark datasets. 
In addition, the results presented in~\cite{Al-riyami2018} are based on only two old benchmark datasets, which makes it hard to draw conclusions about the generalisability of ML-based NIDSs more broadly.
Furthermore, the study only includes the supervised learning models and does not include an evaluation of unsupervised learning models.

The~\cite{Apruzzese2022} is the other work in which the cross evaluation of ML-based NIDSs is discussed.
The main focus of this study is proposing a framework for cross evaluation of ML-based NIDSs by creating various combinations of the Benign and Attack classes of one or more datasets for the training and evaluation.
The proposed framework consists of 10 different combinations in which combination number 4 considers generalisation capabilities.
In this scenario while the Benign class of the same dataset is used for the model training and evaluation, the attack classes of different datasets are used for the training and evaluation.
Then they provide the results for this generalisation evaluation, and show that there is a significant performance drop in detecting various attacks.
The study uses the same 4 datasets used in our study for the uniform scenarios but provides the detection ratios for the Botnet, DoS and Other (all the other classes) separately.
While this is the closest study to what we are doing in this paper, using the Benign class of the same dataset in the training and evaluation makes it different from our work that evaluates generalisability over a full new dataset.

Although the two above works partially study the generalisability of ML-based NIDSs, their main focus is a different subject.
As such, they do not investigate different aspects of  generalisability such as the effect of swapping the source and target datasets.
More importantly, since generalisability is not the main focus of these works, they do not investigate the reason behind lack or existence of the generalisability between two datasets.
In addition, the fact that~\cite{Apruzzese2022} uses the Benign class of the same dataset for the training and evaluation, and the importance of the Benign traffic in ML-based NIDSs as shown in~\cite{2021benchmarking}, makes their investigations completely different from the objective of our study.

\section{Datasets}\label{datasets}
Table~\ref{tab: datasets} shows the summary information of the four datasets used in this study, all in NetFlow (NF) format. 
These datasets,  which include NFv2-UNSW-NB15, NFv2-CIC-2018, NFv2-ToN-IoT and NFv2-BoT-IoT~\cite{sarhan2021towards}, are converted from their original formats published as UNSW-NB15~\cite{unsw}, CIC-2018~\cite{cic}, ToN-IoT~\cite{ton} and BoT-IoT~\cite{bot} respectively.
The first version of the NetFlow format datasets (NFv1) with 20 features has been published in~\cite{sarhan2020netflow}, and the second version, that is used in this work and includes 43 features, is discussed in ~\cite{sarhan2021towards}.
The procedure for converting the original format into NetFlow and the labelling of the converted flows are also explained in~\cite{sarhan2021towards}.

\begin{table}[!b]
\scriptsize
\caption{Summary information of classes in the NetFlow datasets studied in this paper}

\label{tab: datasets}
\centering
\begin{tabular}{
|>{\centering\arraybackslash}m{1.9cm}
|>{\centering\arraybackslash}m{1cm}
|>{\centering\arraybackslash}m{1.6cm}
|>{\centering\arraybackslash}m{0.95cm}
|>{\centering\arraybackslash}m{0.85cm}|
}

\hline
\centering{\multirow{2}{*}{\textbf{Dataset}}}  & 
\multirow{2}{*}{\textbf{Records (\#)}}  & 
\multirow{2}{*}{\textbf{Class}} &
\multirow{2}{*}{\textbf{Class (\#)}} & 
\multirow{2}{*}{\textbf{Class (\%)}} 
\\
& & & &
\\ 
\hline 
\hline

\multirow{4}{*}{\vspace{-.25cm}\textbf{NFv2-BoT-IoT}} & \multirow{4}{*}{\vspace{-.25cm}37,763,497} & Benign & 135,037 & 0.36 \\ \cline{3-5}
 &  & DDoS & 18,331,847 & 48.54 \\ \cline{3-5}
 &  & DoS  & 16,673,183 & 44.15 \\ \cline{3-5}
 &  & Reconnaissance & 2,620,999 & 6.94 \\ \cline{3-5}
 &  & Theft & 2,431 & 0.01 \\ 
 \hline
 \hline

\multirow{15}{*}{\vspace{-1cm}\textbf{NFv2-CIC-2018}} & \multirow{15}{*}{\vspace{-1cm}18,893,708} & Benign  & 16,635,567  &  88.05 \\ \cline{3-5}
 &  & DDOS-HOIC  & 1,080,858 & 5.72 \\ \cline{3-5}
 &  & DoS-Hulk &  432,648 & 2.29 \\ \cline{3-5}
 &  & \tiny{DDoS-LOIC-HTTP}  & 307,300 & 1.63 \\ \cline{3-5}
 &  & Bot & 143,097 & 0.76 \\ \cline{3-5}
 &  & Infilteration & 116,361 & 0.62 \\ \cline{3-5} 
 &  & SSH-Bruteforce   & 94,979 & 0.50 \\ \cline{3-5}
 &  & \tiny{DoS-GoldenEye}  & 27,723 & 0.15 \\ \cline{3-5}
 &  & \tiny{FTP-BruteForce}  & 25,933 & 0.14 \\ \cline{3-5}
 &  & \tiny{DoS-SlowHTTPTest}   & 14,116 & 0.08 \\ \cline{3-5}
 &  & DoS-Slowloris & 9,512 & 0.05 \\ \cline{3-5}
 &  & \tiny{Brute Force-Web}   & 2,143 & 0.01 \\ \cline{3-5}
 &  & \tiny{DDOS-LOIC-UDP} &  2,112 & 0.01 \\ \cline{3-5}
 &  & \tiny{Brute Force-XSS}  & 927 & 0.01 \\ \cline{3-5}
 &  & SQL Injection & 432 & 0.01 \\ 
 \hline
 \hline

\multirow{5}{*}{\vspace{-1.75cm}\textbf{NFv2-ToN-IoT}} & \multirow{5}{*}{\vspace{-1.75cm}16,940,496}  & Benign & 6,099,469 & 36.01 \\ \cline{3-5}
 &  & scanning  & 3,781,419 & 22.32 \\ \cline{3-5}
 &  & xss & 2,455,020 & 14.49 \\ \cline{3-5}
 &  & ddos & 2,026,234 & 11.96 \\ \cline{3-5}
 &  & password & 1,153,323 & 6.81 \\ \cline{3-5}
 &  & dos & 712,609 & 4.21 \\ \cline{3-5}
 &  & injection  & 684,465 & 4.04 \\ \cline{3-5}
 &  & backdoor & 16,809 & 0.1 \\ \cline{3-5}
 &  & mitm & 7,723 & 0.05 \\ \cline{3-5}
 &  & ransomware  & 3,425 & 0.02 \\ 
 \hline
 \hline

\multirow{5}{*}{\vspace{-1cm}\textbf{NFv2-UNSW-NB15}} & \multirow{5}{*}{\vspace{-1cm}2,390,275} & Benign & 2,295,222 & 96.02 \\  \cline{3-5}
 &  & Exploits  & 31,551 & 1.32 \\ \cline{3-5}
 &  & Fuzzers   & 22,310 & 0.93 \\ \cline{3-5}
 &  & Generic  & 16,560 & 0.69 \\ \cline{3-5}
 &  & Reconnaissance   & 12,779 & 0.53 \\ \cline{3-5}
 &  & DoS  & 5,794 & 0.24 \\ \cline{3-5}
 &  & Analysis &  2,299 & 0.10 \\ \cline{3-5}
 &  & Backdoor &  2,169 & 0.09 \\ \cline{3-5}
 &  & Shellcode &  1,427 & 0.06 \\ \cline{3-5}
 &  & Worms & 164 & 0.01 \\ 
 \hline
\end{tabular}
\end{table}

The three original datasets UNSW-NB15, ToN-IoT and BoT-IoT are published by same research group. Since the network setup used for the traffic generation for each dataset is very different, they represent different network environments and represent a valid basis for the evaluation of the generalisability of NIDSs.
This is also the case for CIC-2018, the fourth considered dataset, which has been generated by a different research group in a completely different network setup.

Since the sizes of these datasets are different, e.g. NFv2-UNSW-NB15 has 2,390,275 flows and the BoT-IoT dataset has 37,763,497 flows, we used a stratified (with regards to classes) sampling strategy with a size of 1,000,000 flows and we ran all the experiments on the sampled datasets.  

As can be seen, all these datasets come with a Benign class and various numbers of different attack classes. 
Since the set of attack classes is different for each dataset, we therefore focus on binary classification only, i.e. Benign vs Attack traffic  in our analysis.
Consequently, all attack classes of each dataset are aggregated under a single class called Attack. Hence, the datasets used in this study have two classes, i.e. Benign and Attack.

Initially, when we ran the experiments related to unsupervised learning algorithms, we noticed that the performance (F1-Score) of the models was very poor when trained or evaluated on two datasets NFv2-BoT-IoT and NFv2-ToN-IoT, in both single domain or multi-domain evaluations.
This was mainly due to the unrealistically high imbalance of the datasets, i.e. the ratio of the Attack to Benign records, as can be seen in Table~\ref{tab: datasets}. 
%
Accordingly, we created and used a balanced version of all four datasets (via down sampling) in terms of Attack-Benign labels, and used for the unsupervised learning models experiments. These datasets, extended with a ``-b" suffix such as NFv2-BoT-IoT-b, indicate the balanced version of the (e.g. NFv2-BoT-IoT) dataset.

\section{Single Domain Evaluation}\label{same dataset Evaluation}
In this section we evaluate the ML-based NIDSs using the common method of NIDS evaluation i.e., using a single dataset for the training and evaluation. 
In this method, a publicly available benchmark NIDS dataset is divided into the training and test subsets, and the ML model is trained and tested against these subsets of the same dataset.
In some cases, such as ~\cite{sarhan2020netflow} and \cite{egraphsage}, more than one dataset is selected for the evaluation. However, the ML model is trained and tested/evaluated against each dataset separately, and the generalisability across datasets is not considered. Since the training and test data are both collected from the same environment, this approach is referred to as \textit{single domain evaluation}.

\subsection{Supervised Learning}
Initially, we evaluate supervised learning methods via the single domain evaluation approach.
There are many previous studies which achieve a very high detection performance using this approach, based on the same benchmark NIDS datasets that we consider in this paper, but in their original (non NetFlow) format, in particular UNSW-NB15~\cite{unsw} and CIC-2018~\cite{cic}. 
Since the NetFlow version of these datasets have been published relatively recently, there are only a few studies such as ~\cite{sarhan2020netflow},  \cite{egraphsage} and \cite{niloo-nids}, that have used them. 
However, we cannot use their results in our comparisons because we need to evaluate the same model later in the multi-domain setup, and these studies do not provide such evaluation. As such, we use the result of our experiment, even for the single-domain evaluation, which might be found in the literature.

\begin{table}[!b]
\scriptsize
  \centering
  \caption{Performance (F1-Score (\%)) of 4 supervised (2 shallow and 2 deep) learning methods when trained and evaluated on the same dataset}
    \begin{tabular}
{
|>{\centering\arraybackslash}m{2.1cm}
|>{\centering\arraybackslash}m{1cm}
|>{\centering\arraybackslash}m{1cm}
|>{\centering\arraybackslash}m{1cm}
|>{\centering\arraybackslash}m{1cm}|
}
\toprule
\textbf{Source/Target} & \textbf{Extra Tree} & \textbf{Random Forest} & \textbf{Feed Forward NN} & \textbf{LSTM} \\
    \midrule
    \textbf{NFv2-BoT-IoT} & 99.82\% & 99.82\% & 99.76\% & 99.92\% \\
    \midrule
    \textbf{NFv2-CIC-2018} & 84.62\% & 95.44\% & 46.27\% & 90.17\% \\
    \midrule
    \textbf{NFv2-ToN-IoT} & 77.63\% & 77.33\% & 93.98\% & 76.38\% \\
    \midrule
    \textbf{NFv2-UNSW-NB15} & 91.73\% & 92.17\% & 90.63\% & 92.82\% \\
    \bottomrule
    \end{tabular}%
  \label{tab:self-eval-f1}%
\end{table}%

\begin{table}[!t]
  \centering
  \caption{Performance (F1-Score (\%)) of 3 unsupervised (semi-supervised) learning methods when trained and evaluated on the same dataset}
    \begin{tabular}{|c|c|c|c|}
    \toprule
    \textbf{Source/Target} & \textbf{IsolationForest} & \textbf{oSVM} & \textbf{SGD-oSVM} \\
    \midrule
    \textbf{NFv2-BoT-IoT-b} & 87.58\% & 72.62\% & 72.70\% \\
    \midrule
    \textbf{NFv2-CIC-2018-b} & 85.05\% & 65.62\% & 65.60\% \\
    \midrule
    \textbf{NFv2-ToN-IoT-b} & 57.15\% & 54.65\% & 54.64\% \\
    \midrule
    \textbf{NFv2-UNSW-NB15-b} & 73.67\% & 76.26\% & 76.25\% \\
    \bottomrule
    \end{tabular}%
  \label{tab:self-unsupervised}%
\end{table}%

\begin{table*}[!t]
\renewcommand{\arraystretch}{1.5}
  \centering
  \caption{F1-Score (\%) of supervised learning-based NIDSs for a target domain when trained on three different domains}
    \begin{tabular}{|c|c|c|c|c|c|}
    \hline 
    \multirow{2}[4]{*}{\textbf{Target}} & \multirow{2}[4]{*}{\textbf{Source}} & \multicolumn{4}{c|}{\textbf{Classifier}} \\
\cline{3-6}          &       & \multicolumn{1}{p{5em}|}{\textbf{Extra Tree}} & \multicolumn{1}{p{5em}|}{\textbf{Random Forest}} & \multicolumn{1}{p{5em}|}{\textbf{Feed Forward}} & \multicolumn{1}{p{5em}|}{\textbf{LSTM}} \\
\hline
    \multicolumn{1}{|c|}{\multirow{3}[6]{*}{\textbf{NFv2-BoT-IoT}}} & \textbf{NFv2-CIC-2018} & 0.14\% & 12.91\% & 91.83\% & 54.74\% \\
\cline{2-6}          & \textbf{NFv2-ToN-IoT} & 24.42\% & 46.75\% & 0.56\% & 0.04\% \\
\cline{2-6}          & \textbf{NFv2-UNSW-NB15} & 94.83\% & 7.61\% & 71.38\% & 81.78\%  \\
\hline
\hline
    \multicolumn{1}{|c|}{\multirow{3}[6]{*}{\textbf{NFv2-CIC-2018}}} & \textbf{NFv2-BoT-IoT} & 22.32\% & 22.32\% & 22.32\% & 28.15\% \\
\cline{2-6}          & \textbf{NFv2-ToN-IoT} & 44.83\% & 4.50\% & 57.55\% & 21.82\% \\
\cline{2-6}          & \textbf{NFv2-UNSW-NB15} & 17.47\% & 7.70\% & 34.89\% & 14.20\% \\
\hline
\hline
    \multicolumn{1}{|c|}{\multirow{3}[6]{*}{\textbf{NFv2-ToN-IoT}}} & \textbf{NFv2-BoT-IoT} & 85.50\% & 85.50\% & 85.50\% & 83.65\% \\
\cline{2-6}          & \textbf{NFv2-CIC-2018} & 62.25\% & 30.28\% & 69.01\% & 48.03\% \\
\cline{2-6}          & \textbf{NFv2-UNSW-NB15} & 73.39\% & 0.00\% & 3.82\% & 81.40\% \\
\hline
\hline
    \multicolumn{1}{|c|}{\multirow{3}[6]{*}{\textbf{NFv2-UNSW-NB15}}} & \textbf{NFv2-BoT-IoT} & 4.90\% & 4.90\% & 4.90\% & 4.41\% \\
\cline{2-6}          & \textbf{NFv2-CIC-2018} & 0.57\% & 0.84\% & 0.05\% & 9.63\% \\
\cline{2-6}          & \textbf{NFv2-ToN-IoT} & 0.00\% & 0.25\% & 0.00\% & 0.40\% \\
    \hline
    \end{tabular}%
  \label{tab:else-supervised}%
\end{table*}%


We chose four simple supervised learning models including two deep and two shallow learning methods.
For the deep learning we chose the same number of layers and nodes from two different architectures. The first model is a simple Feed Forward neural network with 5 hidden layers, each with 10 nodes. 
The second model is a Long Short-Term Memory (LSTM) network with the same number of layers and nodes on each layer. 
For the shallow learning methods, we used a Random Forest and an Extra-Tree classifier, which allow us to easily implement our NIDSs without much effort to tune hyperparameters.

It is possible to search and fine tune the hyperparameters of the ML-models to achieve the best possible performances on a training dataset.
However, this increases the chance of over-fitting to the training dataset and reduces the performance on other datasets, not seen during the training.

Since in the next steps of this study we are going to evaluate the performance of these models via the multi-domain approach, hyperparameter fine tuning will bias the performance results towards the single domain evaluation.
Accordingly, in order to avoid over-fitting to the training datasets, we used the default hyperparameters for the different models, as provided by the corresponding software libraries, i.e. scikit-learn (for the shallow learning methods)~\cite{scikit-learn} and TensorFlow (for the deep learning methods)~\cite{tensorflow2015-whitepaper}.

Table~\ref{tab:self-eval-f1} shows the classification performance (F1-Score) for the four supervised learning models trained and evaluated on the same NetFlow dataset (single-domain evaluation). 
As can be seen, the performance on each dataset are mostly consistent across all four ML-based NIDSs (1\% to 5\% variations), except the Feed Forward NN model, which its performance is up to 50\% different, in one case, from the rest of models. 
It is noticeable that even these simple models, without fine tuning of their hyperparameters, are able to achieve a very high classification performance in most cases.
For instance, all these simple models have been able to achieve a F1-Score above $99\%$ on the NFv2-BoT-IoT dataset, and for the rest of the datasets there is at least one NIDS model which achieves a F1-Score greater than $92\%$.

\subsection{Unsupervised Learning}\label{SDE-Unsupervised}
While the generalisability of supervised learning models has partly been investigated for the NIDS application in a previous study~\cite{Al-riyami2018}, there are no such results for unsupervised learning methods in the context of NIDS, to the best of our knowledge.
As such, we have included three unsupervised learning models in our investigation. These models, which include the \textit{Isolation Forest (IsolationForest)}~\cite{Liu2012}, \textit{One-Class Support Vector Machines(oSVM)}~\cite{oSVM2013}, and \textit{Stochastic Gradient Descent one-Class Support Vector Machines (SGD-oSVM)}~\cite{SGD-oSVM2004}, are used for the purpose of anomaly/attack detection in our context.

Although these models are typically considered unsupervised learning algorithms, since they are very sensitive to anomalies/outliers in the training set~\cite{oSVM2013}, we have used them in a semi-supervised manner.
This means the models are exposed to one class of data during the training phase, but are tested against both classes. 

To achieve this, each benchmark dataset is divided into the training and test subsets. Then, the Attack samples are removed from the training subset.
Hence, the models are trained only on normal/Benign data samples and evaluated against the test subset, which includes samples of both the Benign and Attack classes. 
%

Similar to the case of the supervised learning models, we did not fine tune the hyperparameters of the models, for achieving the best possible performances, and mostly used the default values provided in the scikit-Learn package.
Table~\ref{tab:self-unsupervised} shows the results of applying these three NIDS models on the four considered benchmark datasets.
The main conclusion from this table can be summarised in two points. First, all models have lower performance on NFv2-ToN-IoT, which was also the case for the supervised learning models (except the Feed Forward model).
Second, the Isolation Forest model has higher performances across all the four datasets.


\begin{figure*}[!t]
    \centering
    \newcommand{\x}{0.8\columnwidth}
    \subfloat[\centering ][Extra Tree]
    {\includegraphics[width=\x]{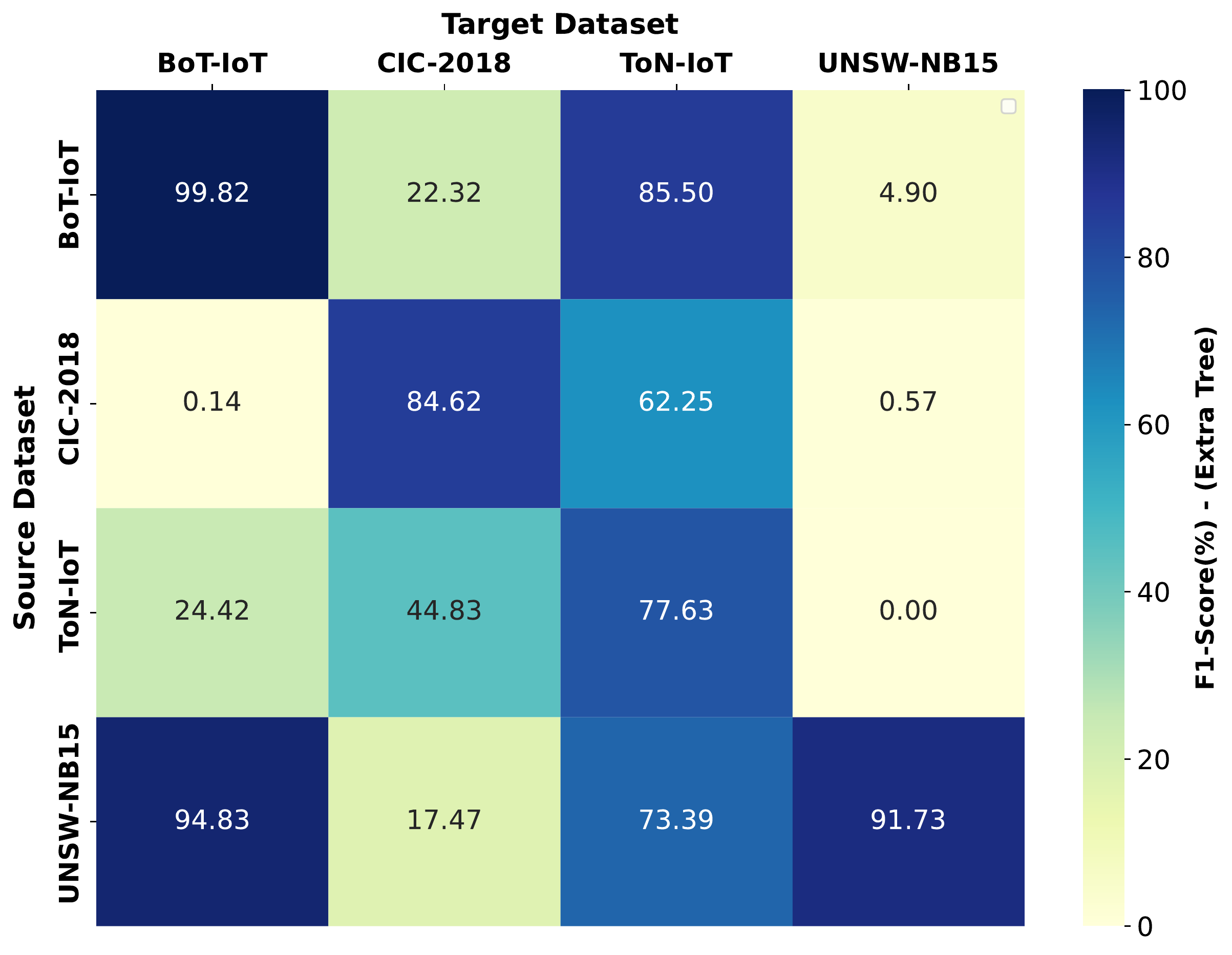}}%
    \hspace{0.75cm}
    \subfloat[\centering ][Random Forest]
    {\includegraphics[width=\x]{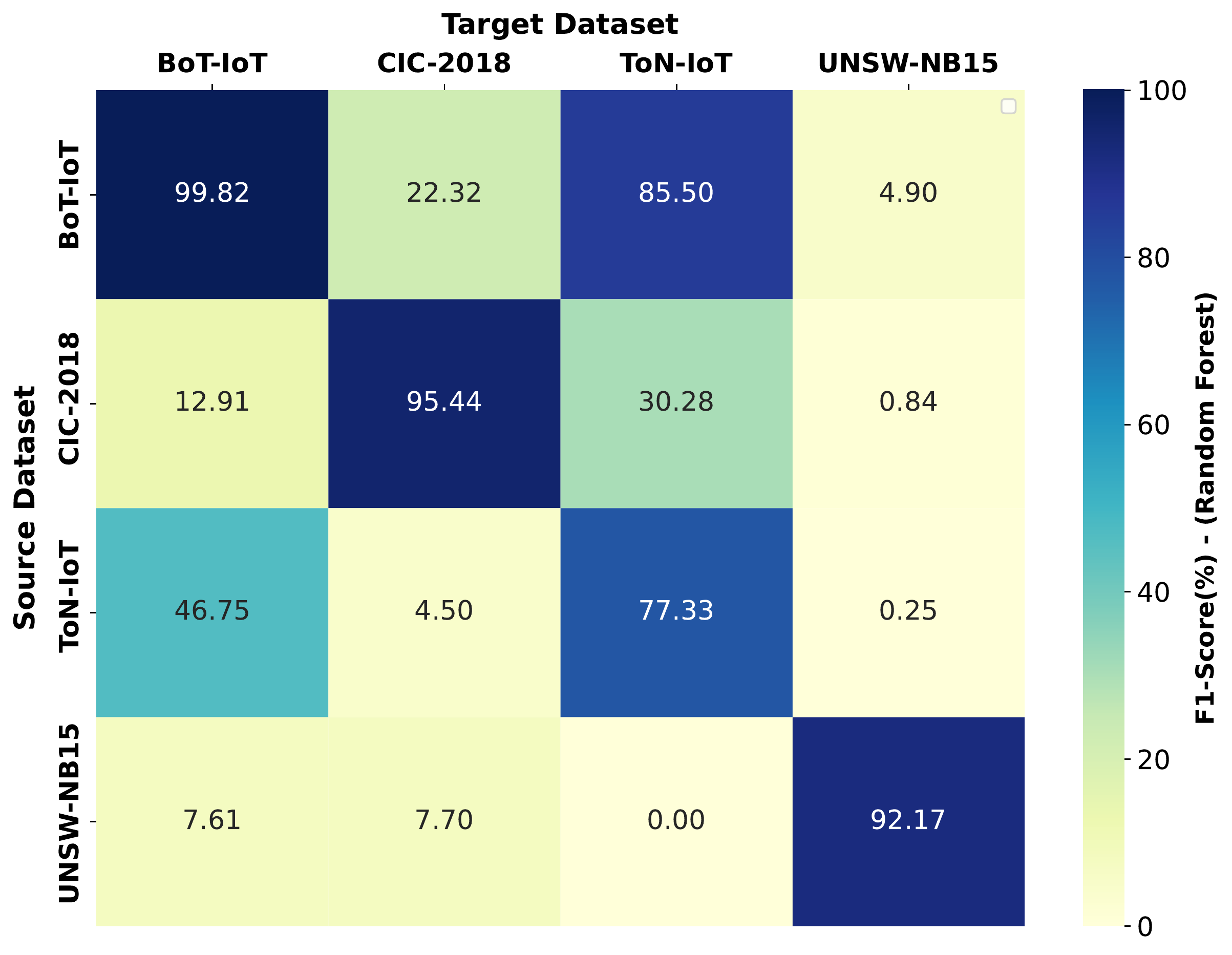}}%
    \vspace{.5cm}
    \qquad
    \subfloat[\centering ][Feed Forward]
    {\includegraphics[width=\x]{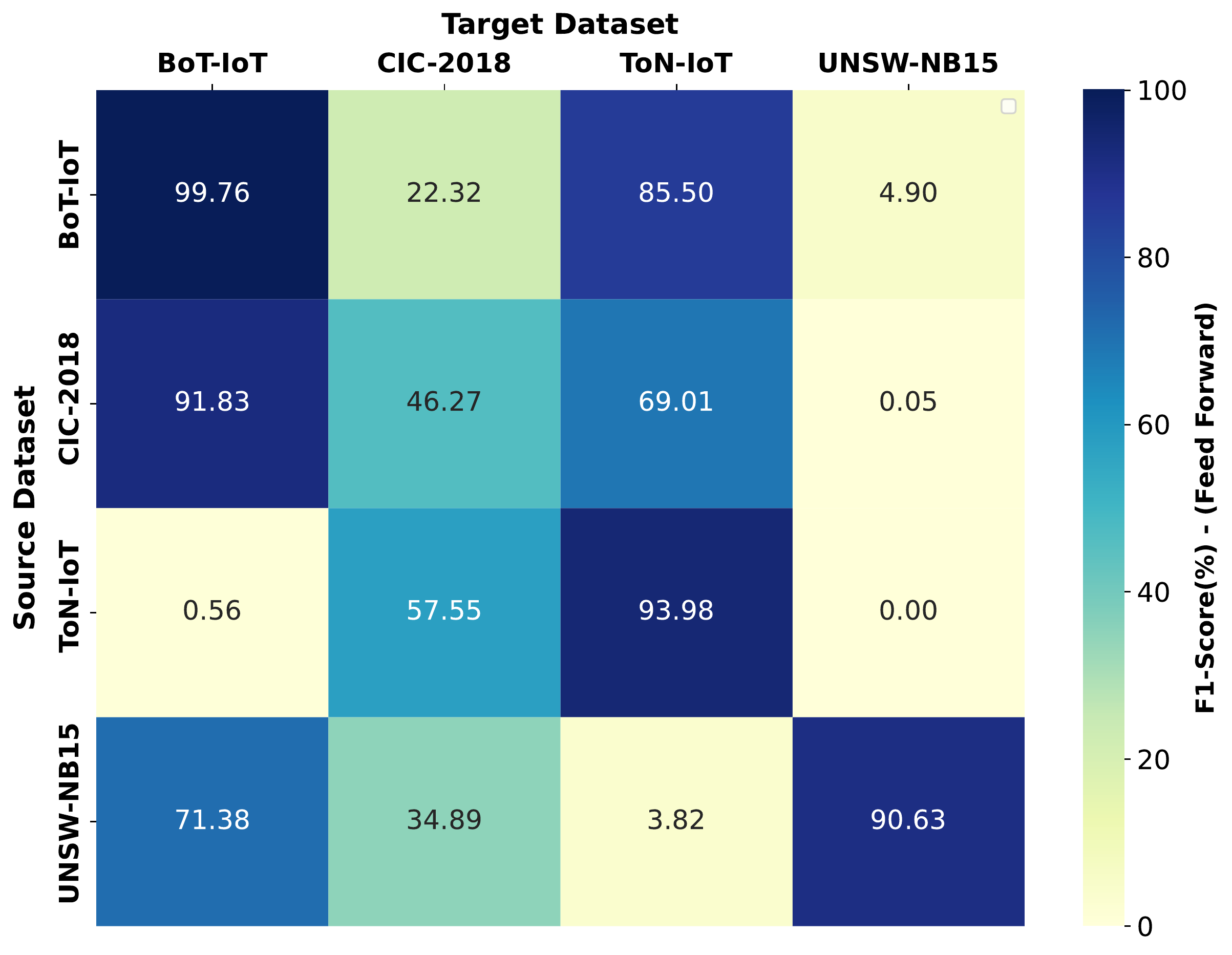}}%
    \hspace{0.75cm}
    \subfloat[\centering ][LSTM]
    {\includegraphics[width=\x]{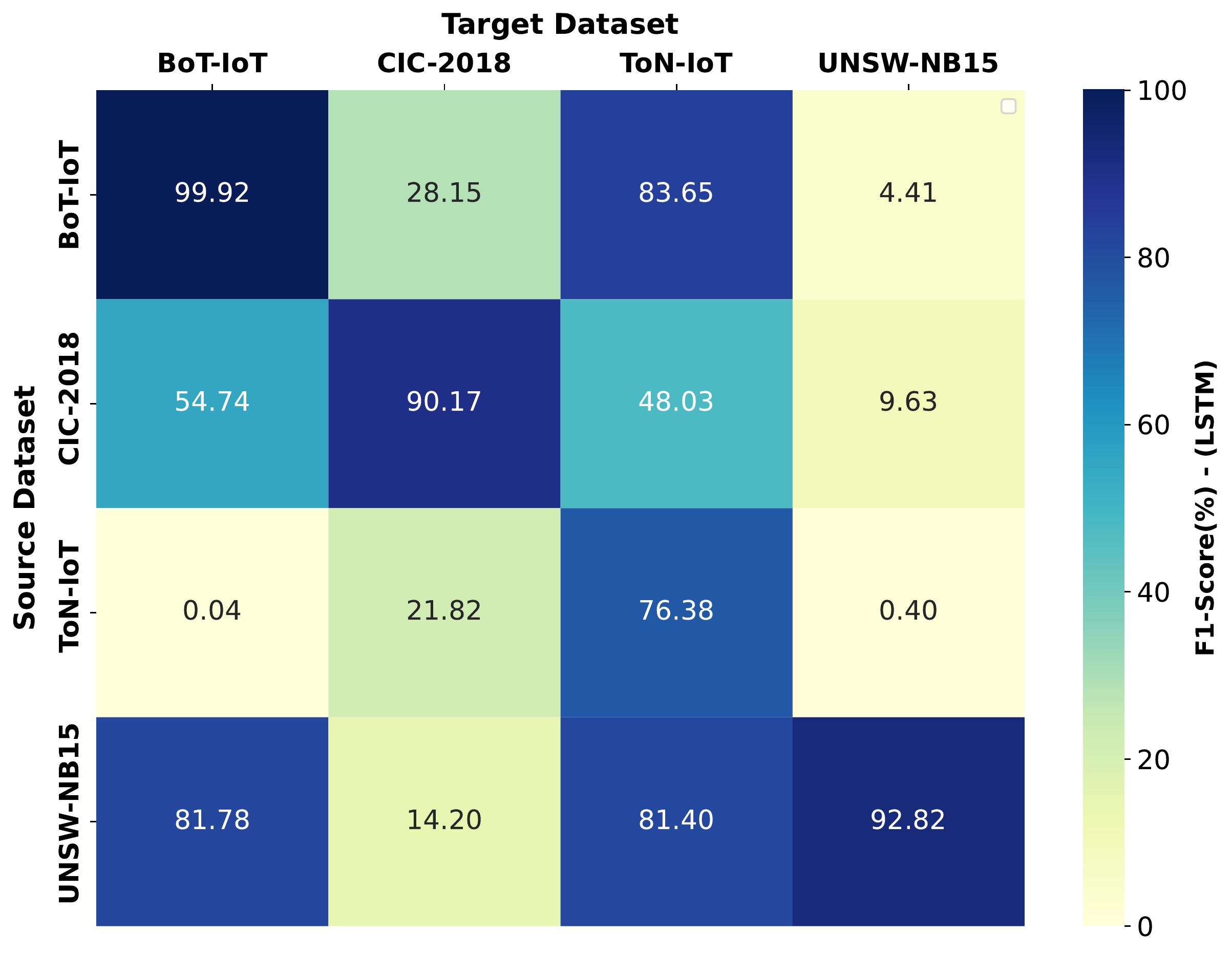}}%
    \caption{F1-Score (\%) of supervised learning-based NIDSs when trained and tested on different datasets for (a) Extra Tree, (b) Random Forest, (c) Feed Forward, and (d) LSTM models. The diagonal entries show the single domain evaluation and off-diagonal values indicate the generalisability evaluation.}%
    \label{fig:SE-trgt-summary}
\end{figure*}

\section{Generalisability Evaluation}\label{Generalisability Evaluation}
In the previous section we evaluated the performance of both supervised and unsupervised learning NIDS algorithms via a single domain approach, which is the current de-facto standard in the NIDS literature.
As it was shown, we were able to get close to the high performance values reported in the literature, using our simple non-fine-tuned models, for most of the cases in both the supervised and unsupervised learning methods.

While having a high performance in a single domain evaluation is a necessary condition for any NIDS targeting the real-world application, there are other conditions to be met as well.
One such condition is the generalisability, i.e., the ability to generalise and translate the high performance on a single domain to other domains.

%

In order to evaluate the generalisability of an NIDS model, we need to separate domains/datasets. The first domain/dataset, which is called the \textit{source domain}, is used for the training, and the second domain/dataset, which is referred to as the \textit{target domain}, is used for the evaluation. 
In this way, the NIDS model is only trained on the source domain and it does not see the target domain during training.

\subsection{Supervised Learning}
Here we evaluate the generalisability of the same four supervised learning models considered in the previous section.
We ran a set of four experiments for each NIDS model in which the model is trained on a source benchmark dataset and evaluated against the other three benchmark datasets without any further training. 
Hence, for each target domain/dataset we have three different results for the same NIDS model, each indicating the performance of the model when trained on a different source domain/dataset.

Table~\ref{tab:else-supervised} shows the results of these experiments, ordered by the target domain and source domain, to allow easy comparison with the single domain results shown in Table~\ref{tab:self-eval-f1}.
As can be seen, each supervised learning-based NIDS model is evaluated for the 12 different source/target domain combinations. 
%

%

Figure~\ref{fig:SE-trgt-summary} provides a more visual representation of both the single-domain and the multi-domain evaluation results, with each sub-figure showing the F1-Score results for a different supervised model. 
The used colour map indicates a very high F1-Score (100\%) in dark blue, and a very low value (0\%) in light yellow, with the in-between values as indicated.   
The source datasets are indicated on the vertical axis, and the corresponding target datasets are shown on the horizontal axis. 
In order to make the name of dataset readable in this figure, we used larger font sizes, that necessitated to remove the ``NFv2-" prefix from the name of all datasets to make them fit to the spaces.
On the diagonal, we observe the single-domain results, where a model is trained and evaluated on the same dataset. The off-diagonal results  show the multi-domain results, which provide information about the degree of generalisability of the different models.

%

The mostly dark colouring on the diagonal indicates a generally high performance in the single-domain evaluation of all the four supervised models.
%
However, the mostly lighter colours, and hence lower F1-Scores, indicate a generally poor ability of the models to generalise from a source dataset to a different target dataset. There are some exceptions though. For example, for the Extra Tree model (Figure~\ref{fig:SE-trgt-summary}-a), UNSW-NB15 as the source domain generalises well to the BoT-IoT dataset as the target domain, with an F1-Score of 94.83\%. 

Interestingly, this result is highly asymmetrical. If we swap the source and target domain, and use BoT-IoT as the source and UNSW-NB15 as the target, the Extra Tree only achieves 4.90\%.
While this is the most prominent example, we observe a generally high degree of asymmetry of the generalisability across different source/target domain pairs and supervised learning models. 

If we compare the generalisability results across the four different ML models, we observe some consistent patterns, but we also notice some significant differences. The Random Forest model seems to perform quite differently from the other three modeles, in particluar for the UNSW-NB15 dataset as the source domain (bottom row). 

Finally, we also observe significant differences among the datasets. Most strikingly, we see that if the UNSW-NB15 dataset is chosen as the target domain, the results are very poor for any of the other datasets chosen as the source domain (rightmost column). This is consistent across all four ML models. 







Table~\ref{tab:supervised-decay} shows the average performance decay per model when evaluated on a dataset not used for training versus datasets used for the training.
Each column indicates a supervised learning model and each row indicates a source/training dataset. 
For instance, the cell in the first row and first column shows the average decay of the performance of the Extra Tree model trained on NFv2-BoT-IoT when tested on NFv2-CIC-2018, NFv2-ToN-IoT and NFv2-UNSW-NB15 datasets, compared to its performance when tested on NFv2-BoT-IoT dataset.
Investigating the performance decays presented in
Table~\ref{tab:supervised-decay} indicates that: 
\\
\begin{table}[!t]
\renewcommand{\arraystretch}{1.5}
\scriptsize
  \centering
  \caption{Average performance (F1-Score (\%)) decay per model and source domain for the supervised learning models}
    \begin{tabular}
    {
|>{\centering\arraybackslash}m{1.5cm}
|>{\centering\arraybackslash}m{0.8cm}
|>{\centering\arraybackslash}m{0.8cm}
|>{\centering\arraybackslash}m{0.8cm}
|>{\centering\arraybackslash}m{0.8cm}
|>{\centering\arraybackslash}m{1cm}|
} 
    \hline
    \tiny\textbf{Source/Training dataset} & \tiny\textbf{Extra Tree Decay (avg.) } & \tiny\textbf{Random Forest Decay (avg.)} & \tiny\textbf{Feed Forward Decay (avg.)} & \tiny\textbf{LSTM Decay (avg.)} & \tiny\textbf{Average Decay per Source} \\
    \hline
    \tiny\textbf{NFv2-BoT-IoT} & 62.24\% & 62.25\% & 62.19\% & 61.18\% & \textbf{61.96\%} \\
    \hline
    \tiny\textbf{NFv2-CIC-2018} & 63.63\% & 80.77\% & -7.36\% & 52.70\% & \textbf{47.43\%} \\
    \hline
    \tiny\textbf{NFv2-ToN-IoT} & 54.55\% & 60.16\% & 74.60\% & 68.96\% & \textbf{64.57\%} \\
    \hline
    \tiny\textbf{NFv2-UNSW-NB15} & 29.84\% & 87.07\% & 53.93\% & 33.70\% & \textbf{51.13\%} \\
    \hline
    \tiny\textbf{Average Decay per Model} & \textbf{52.57\%} & \textbf{72.56\%} & \textbf{45.84\%} & \textbf{54.13\%} &  \textbf{56.28\%}\\
    \hline
    \end{tabular}%
  \label{tab:supervised-decay}%
\end{table}%

\begin{enumerate}[(a)]
\item There is no supervised learning-based NIDS generalising over all combination of source-target domains.
\item Deep learning-based NIDSs generalise better than shallow learning-based NIDSs in average. 
\item There is an average performance decay of $56.28\%$ when a supervised learning model is evaluated on a dataset other than its training dataset.
\end{enumerate}

\subsection{Unsupervised Learning}


Similar to the supervised learning algorithms, we evaluated the generalisability of unsupervised learning-based NIDSs using the same four benchmark datasets as we used for the single domain evaluation.
As per the single domain evaluation, training is performed using only the normal/Benign class of the source domain, while the trained model is exposed to both the Benign and Attack classes of the target domain for the evaluation of its generalisability.

\begin{table}[!b]
\renewcommand{\arraystretch}{1.3}
\scriptsize
  \centering
  \caption{F1-Score (\%) of unsupervised learning-based NIDSs for a target domain when trained on three different domains}
    \begin{tabular}{|c|c|c|c|c|}
    \hline
    \textbf{target} & \textbf{source} & \textbf{IsolationForest} & \textbf{oSVM} & \tiny\textbf{SGD-oSVM} \\
    \hline
    \multirow{3}[6]{*}{\tiny\textbf{NFv2-BoT-IoT-b}} & \tiny\textbf{NFv2-CIC-2018-b} & 49.35\% & 77.87\% & 77.87\% \\
\cline{2-5}          & \tiny\textbf{NFv2-ToN-IoT-b} & 60.52\% & 80.38\% & 80.86\% \\
\cline{2-5}          & \tiny\textbf{NFv2-UNSW-NB15-b} & 34.33\% & 0.23\% & 0.23\% \\
    \hline\hline
    \multirow{3}[6]{*}{\tiny\textbf{NFv2-CIC-2018-b}} & \tiny\textbf{NFv2-BoT-IoT-b} & 0.18\% & 56.86\% & 58.71\% \\
\cline{2-5}          & \tiny\textbf{NFv2-ToN-IoT-b} & 0.25\% & 56.24\% & 58.12\% \\
\cline{2-5}          & \tiny\textbf{NFv2-UNSW-NB15-b} & 13.84\% & 56.66\% & 57.74\% \\
    \hline\hline
    \multirow{3}[6]{*}{\tiny\textbf{NFv2-ToN-IoT-b}} & \tiny\textbf{NFv2-BoT-IoT-b} & 66.75\% & 29.20\% & 30.45\% \\
\cline{2-5}          & \tiny\textbf{NFv2-CIC-2018-b} & 62.12\% & 22.23\% & 22.17\% \\
\cline{2-5}          & \tiny\textbf{NFv2-UNSW-NB15-b} & 0.00\% & 7.10\% & 7.10\% \\
    \hline\hline
    \multirow{3}[6]{*}{\tiny\textbf{NFv2-UNSW-NB15-b}} & \tiny\textbf{NFv2-BoT-IoT-b} & 0.00\% & 76.29\% & 76.28\% \\
\cline{2-5}          & \tiny\textbf{NFv2-CIC-2018-b} & 26.22\% & 64.80\% & 64.80\% \\
\cline{2-5}          & \tiny\textbf{NFv2-ToN-IoT-b} & 0.00\% & 64.85\% & 64.85\% \\
    \hline
    \end{tabular}%
  \label{tab:else-unsupervised}%
\end{table}%

\begin{figure}[!t]
    \centering
    \newcommand{\x}{0.8\columnwidth}
    \subfloat[\centering ][Isolation Forest]
    {\includegraphics[width=\x]{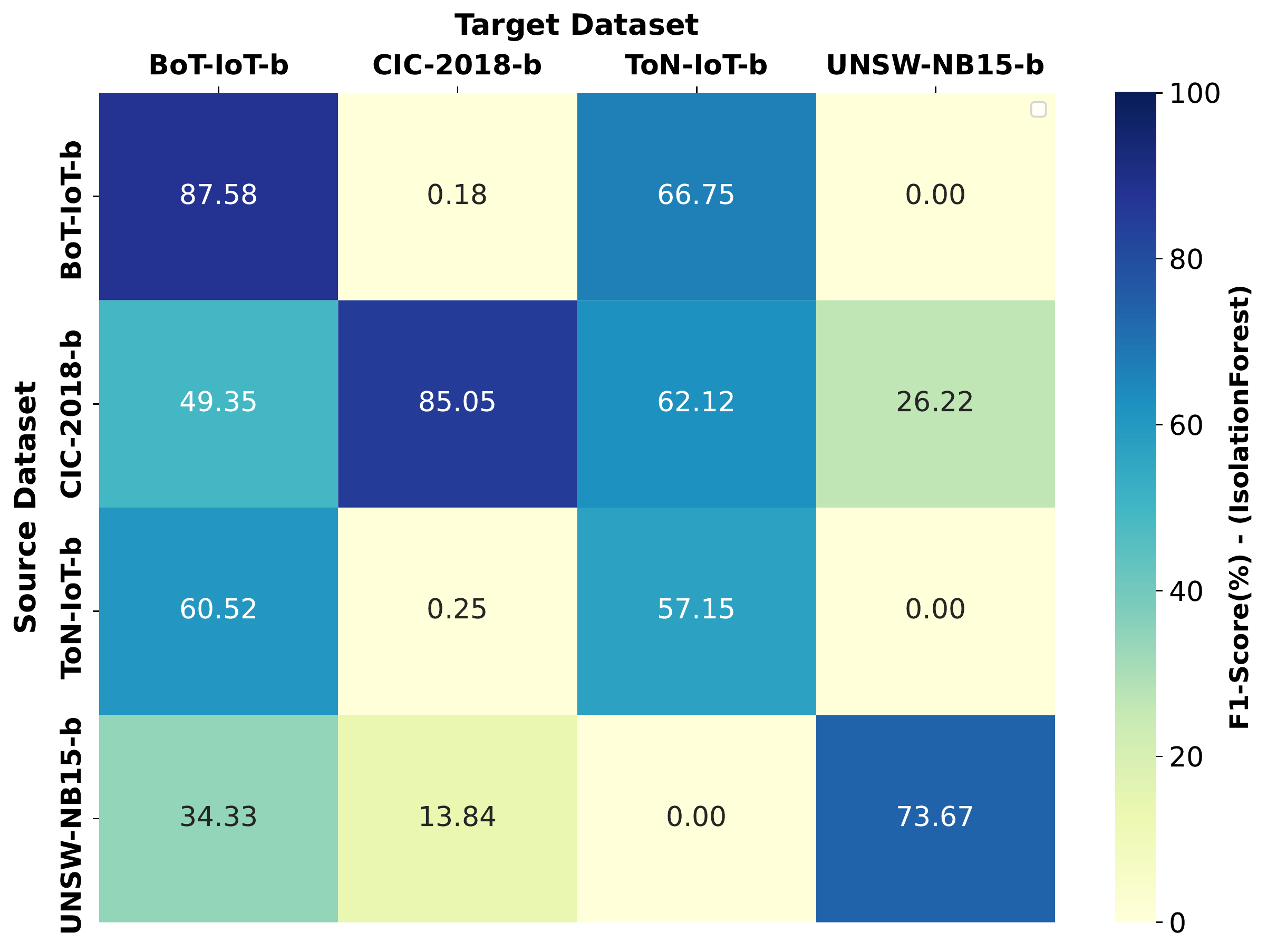}}%
    \vspace{0.4cm}
    \subfloat[\centering ][oSVM]
    {\includegraphics[width=\x]{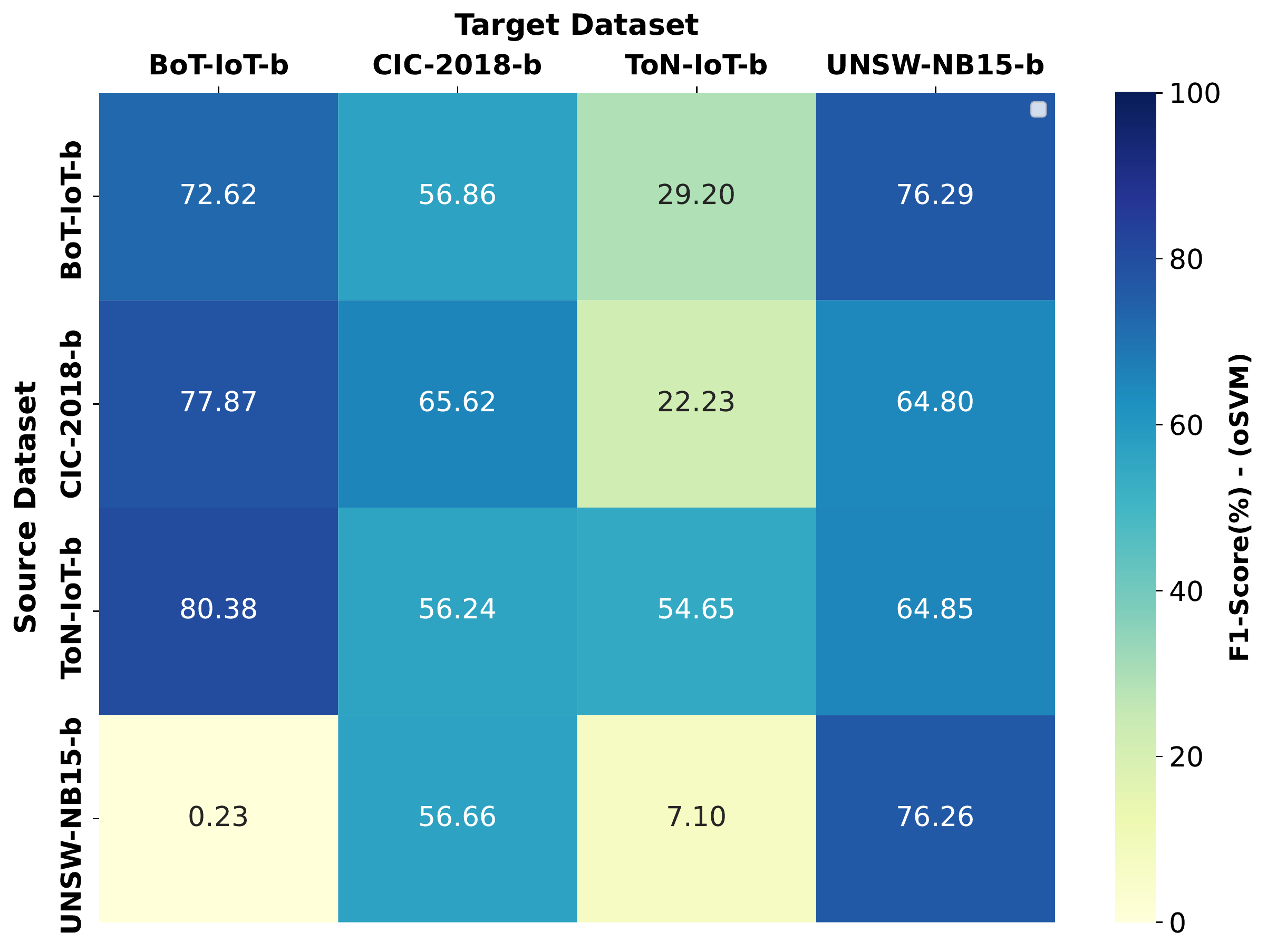}}%
    \vspace{.4cm}
    \qquad
    \subfloat[\centering ][SGD-oSVM]
    {\includegraphics[width=\x ]{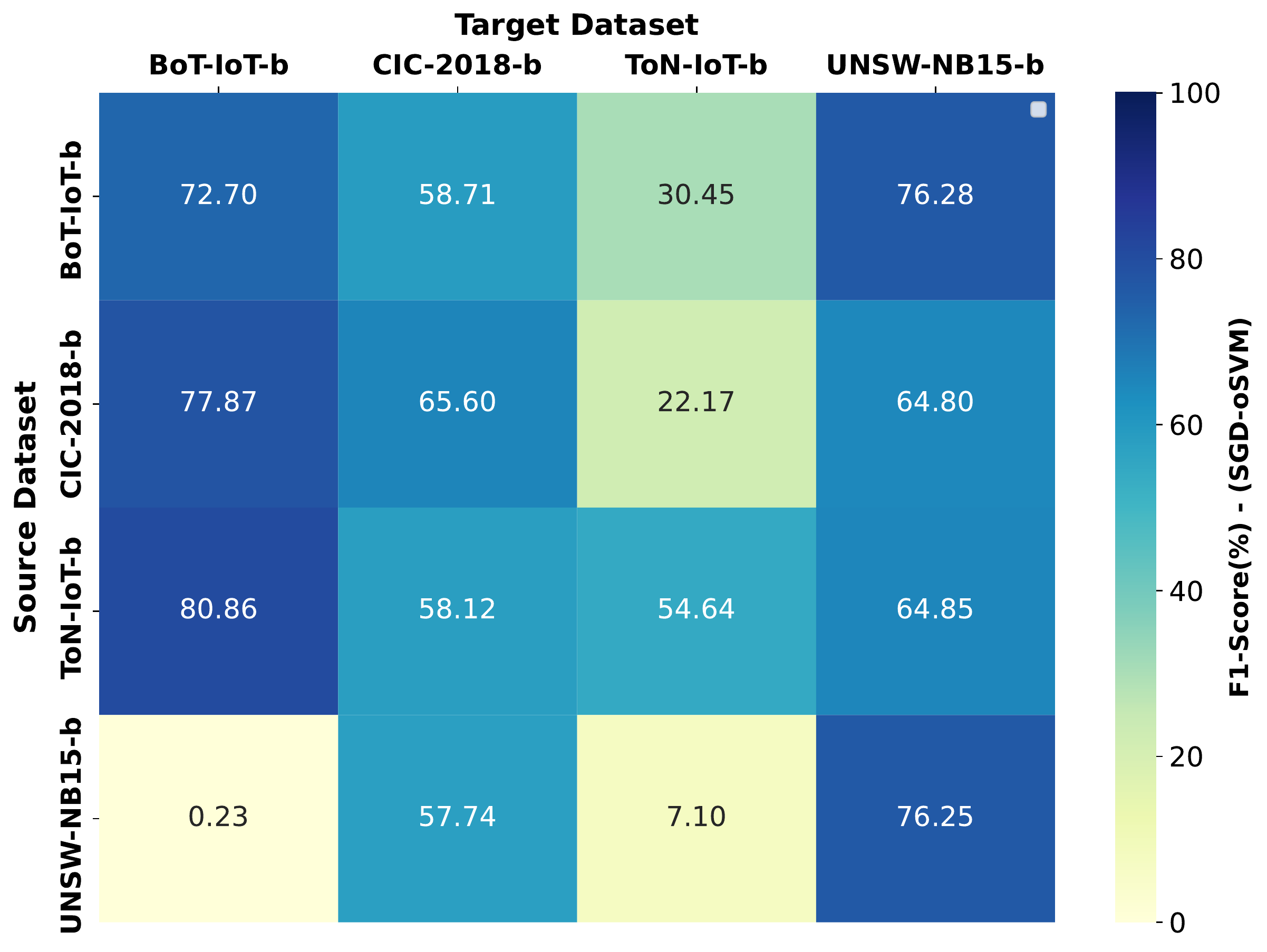}}%
    \caption{F1-Score (\%) of unsupervised learning-based NIDSs when trained and tested on different datasets for (a) Isolation Forest, (b) oSVM, and (c) SGD-oSVM models. The diagonal entries show the single domain evaluation and off-diagonal values indicate the generalisability evaluation.}%
    \label{fig:U-SE-trgt-summary}
\end{figure}


Table~\ref{tab:else-unsupervised} shows the results of generalisability evaluation of the unsupervised learning algorithms.
Similar to supervised learning algorithms, each model is trained on one dataset and evaluated on the other three datasets that were not used for training. 
For each of the four datasets used as the source domain, we consider the other three as the target domain, resulting in 12 target/source domain combinations that are the basis for the generalisability evaluation. 

%

Figure~\ref{fig:U-SE-trgt-summary} visualises the results for the three considered unsupervised algorithms, and also includes the single-domain results on the diagonal, corresponding to Figure~\ref{fig:SE-trgt-summary}.
%
%
As can be seen, while the performance of the models is not as high as the supervised learning algorithms, in the single domain experiments, there are many cases in which the generalisability is equal or higher than single domain evaluation.


Table~\ref{tab:unsupervised-decay} quantifies these outcomes in terms of the performance (F1-Score) decays.
It shows the average performance decay for each unsupervised learning model when evaluated on three datasets not used for the training compared to evaluation on the training dataset.
Each column indicates an unsupervised learning model, each row indicates a source domain, and each value indicates the average of three F1-Score values. 
Investigating these performance decays indicates that

\begin{enumerate}[(a)]
\item There is no unsupervised learning model that generalises over all combination of source-target domains.
\item The two unsupervised algorithms oSVM and SGD-oSVM are better generalised than Isolation Forest, even though their single domain performance is weaker.
\item There is an average performance decay of $28.33\%$ when an unsupervised learning model is evaluated on a dataset other than its training dataset.
\end{enumerate}

\subsection{Supervised vs Unsupervised Learning Models:}\label{svg}
Comparing the results shown in  Table~\ref{tab:supervised-decay} and Table~\ref{tab:unsupervised-decay} indicates 
\begin{enumerate}[(a)]
\item The unsupervised learning models generalise better with an overall average performance decay of $28.33\%$, compared to the supervised learning models with an overall average performance decay of $56.28\%$.
\item The superiority of unsupervised learning models is consistent across all source domains, i.e., the average per source domain decay (last column of Tables~\ref{tab:supervised-decay} and ~\ref{tab:unsupervised-decay}) is significantly lower for unsupervised learning models for all the four datasets.    
\end{enumerate}


\begin{figure*}[!t]
\newcommand{\x}{4.5cm}
    \subfloat[\centering ][Source: BoT-IoT, Target: BoT-IoT]
    {\includegraphics[width=0.95\columnwidth, height=\x]{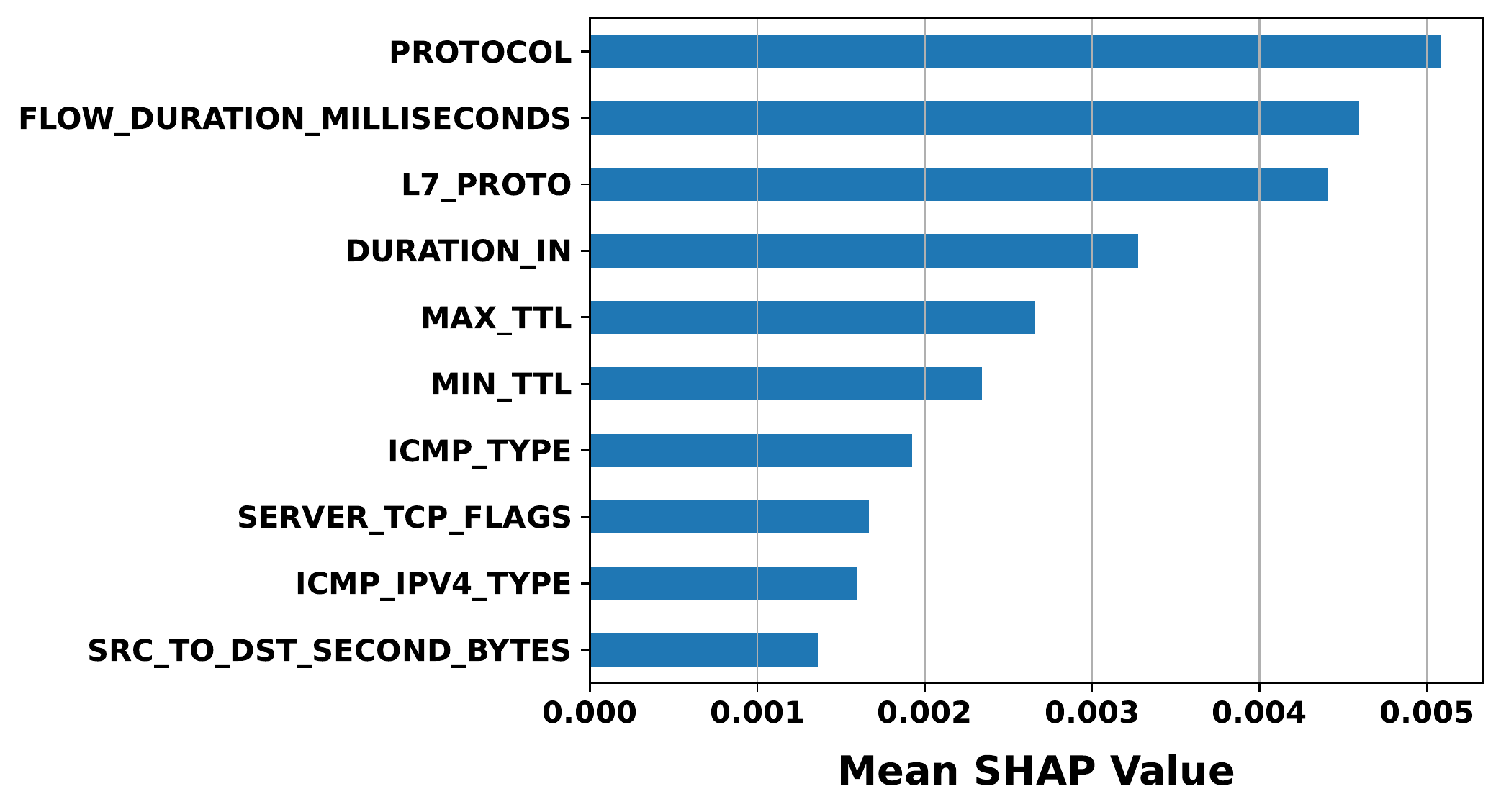}}%
    \hspace{1cm}
    \subfloat[\centering ][Source: BoT-IoT, Target: UNSW-NB15]
        {\includegraphics[width=0.95\columnwidth, height=\x]{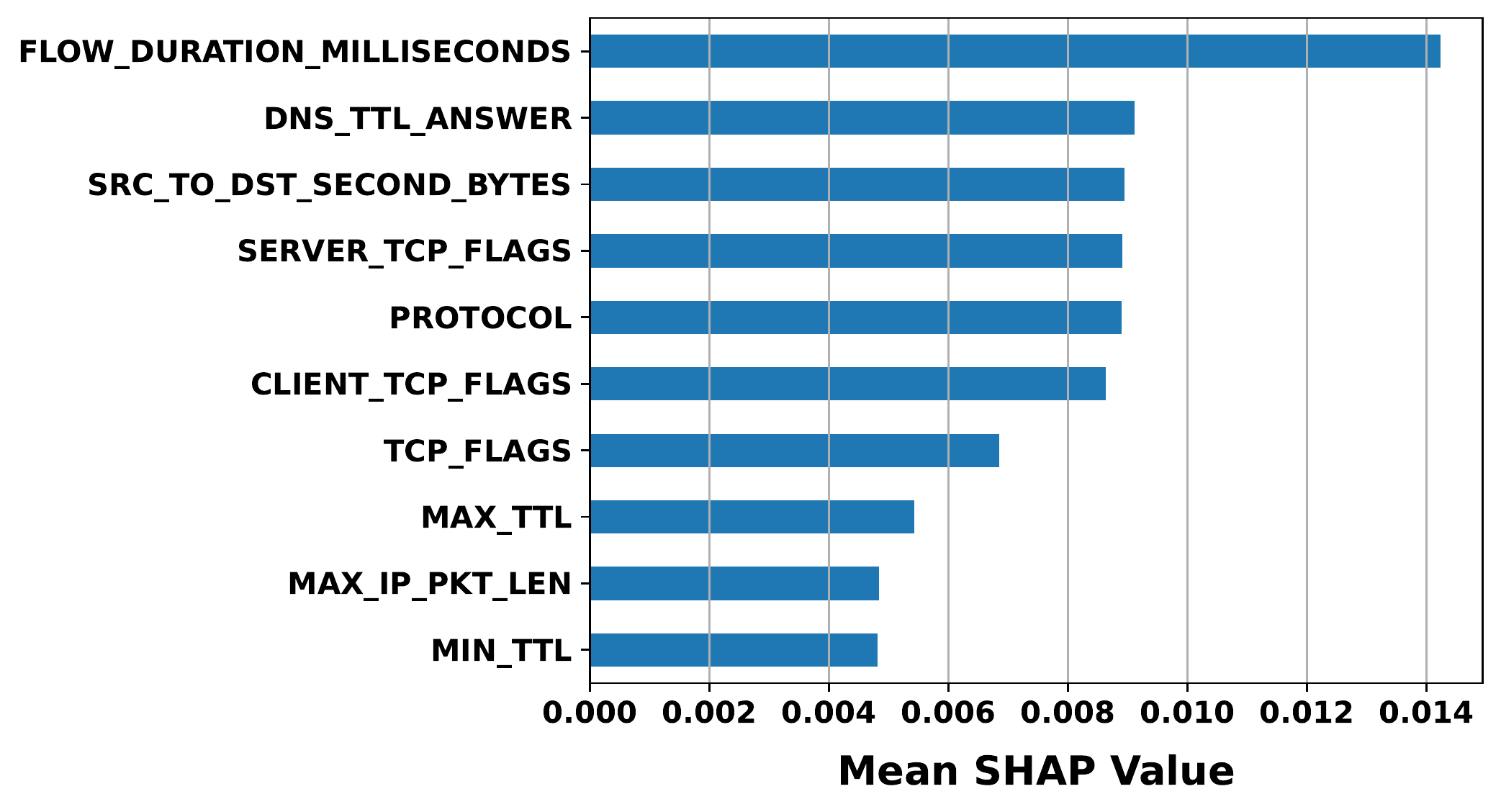}}%
    \caption{Feature importance (Mean absolute SHAP value) of top 10 features for the Feed Forward model trained on NFv2-BoT-IoT dataset and evaluated on(a) NFv2-BoT-IoT and (b) NFv2-UNSW-NB15 datasets}%
    \label{fig:shap-mean}
\end{figure*}

\begin{table}[!t]
\renewcommand{\arraystretch}{1.5}
\scriptsize
  \centering
  \caption{Average performance (F1-Score (\%)) decay per model and source for the unsupervised learning models}
    \begin{tabular}
            {
|>{\centering\arraybackslash}m{2.2cm}
|>{\centering\arraybackslash}m{1.2cm}
|>{\centering\arraybackslash}m{0.9cm}
|>{\centering\arraybackslash}m{1cm}
|>{\centering\arraybackslash}m{1cm}|
} 
    \hline
    \tiny\textbf{Source/Training dataset} & \tiny\textbf{Isolation Forest Decay (avg.)} & \tiny\textbf{oSVM Decay (avg.)} & \tiny\textbf{SGD-oSVM Decay (avg.)} & \tiny\textbf{Decay per source (avg.)} \\
    \hline
    \textbf{NFv2-BoT-IoT-b} & 39.51\% & 19.80\% & 19.71\% & \textbf{26.34\%} \\
    \hline
    \textbf{NFv2-CIC-2018-b} & 80.29\% & 9.04\%  & 7.41\%  & \textbf{32.25\%} \\
    \hline
    \textbf{NFv2-ToN-IoT-b} & 14.19\% & 35.14\% & 34.73\% & \textbf{28.02\%} \\
    \hline
    \tiny{\textbf{NFv2-UNSW-NB15-b}} & 64.93\% & 7.61\%  & 7.60\%  & \textbf{26.72\%} \\
    \hline
    \textbf{Decay per Model (avg.) } & \textbf{49.73\%} & \textbf{17.90\%} & \textbf{17.36\%} & \textbf{28.33\%} \\
    \hline
    \end{tabular}%
   \label{tab:unsupervised-decay}%
\end{table}%

\section{ Explaining Generalisability}
\label{Analysis} 
%
Explaining the behavior of a machine learning model usually requires to investigate the impact of the features of input data on output.
There are a range of tools and techniques to study and estimate the feature importance and how much each feature impacts the model output.
SHapley Additive exPlanations (SHAP) values~\cite{shapley2017} is one of the recent trends in explaining and interpreting the output of the AI/ML models in terms of the features of the datasets.  
It provides a value for each feature in the train/test datasets, which indicates how much a feature has contributed to the generated output.
Hence, these values depend on the training dataset, the ML model, and the evaluation dataset.

Figure~\ref{fig:shap-mean} shows the mean absolute SHAP value for the top ten features where the Feed forward model is trained on NFv2-BoT-IoT and evaluated on (a) NFv2-BoT-IoT and (b) NFv2-UNSW-NB15.
These figures, called the feature importance plot, show the features in descending order (of their mean of absolute SHAP values) on the vertical axis and the mean SHAP value on the horizontal axis. 

Comparing the two feature importance plots clearly shows that feature orders and the mean SHAP values of features significantly varies between the two. 
For instance, in Figure~\ref{fig:shap-mean}-(a), PROTOCOL is the most important feature with a mean SHAP value of $0.005$, while in Figure~\ref{fig:shap-mean}-(b), the most important feature is FLOW\_DURATION\_IN\_MILLISECONDS with a mean SHAP value of $0.014$, and PROTOCOL is fifth important feature with a mean SHAP value of $0.008$.
This is a clear indication that the behaviour of the Feed Forward model trained on NFv2-BoT-IoT is entirely different when tested on NFv2-BoT-IoT datasets and when tested on NFv2-UNSW-NB15. 
This conclusion is consistent with the results shown in  Figure~\ref{fig:SE-trgt-summary}-(C) for the Feed Forward model.


\begin{figure}[!t]
\newcommand{\x}{4.5cm}
\newcommand{\y}{0.5cm}
    \centering
    \subfloat[\centering ][Isolation Forest]
    {\includegraphics[width=1\columnwidth, height=\x]{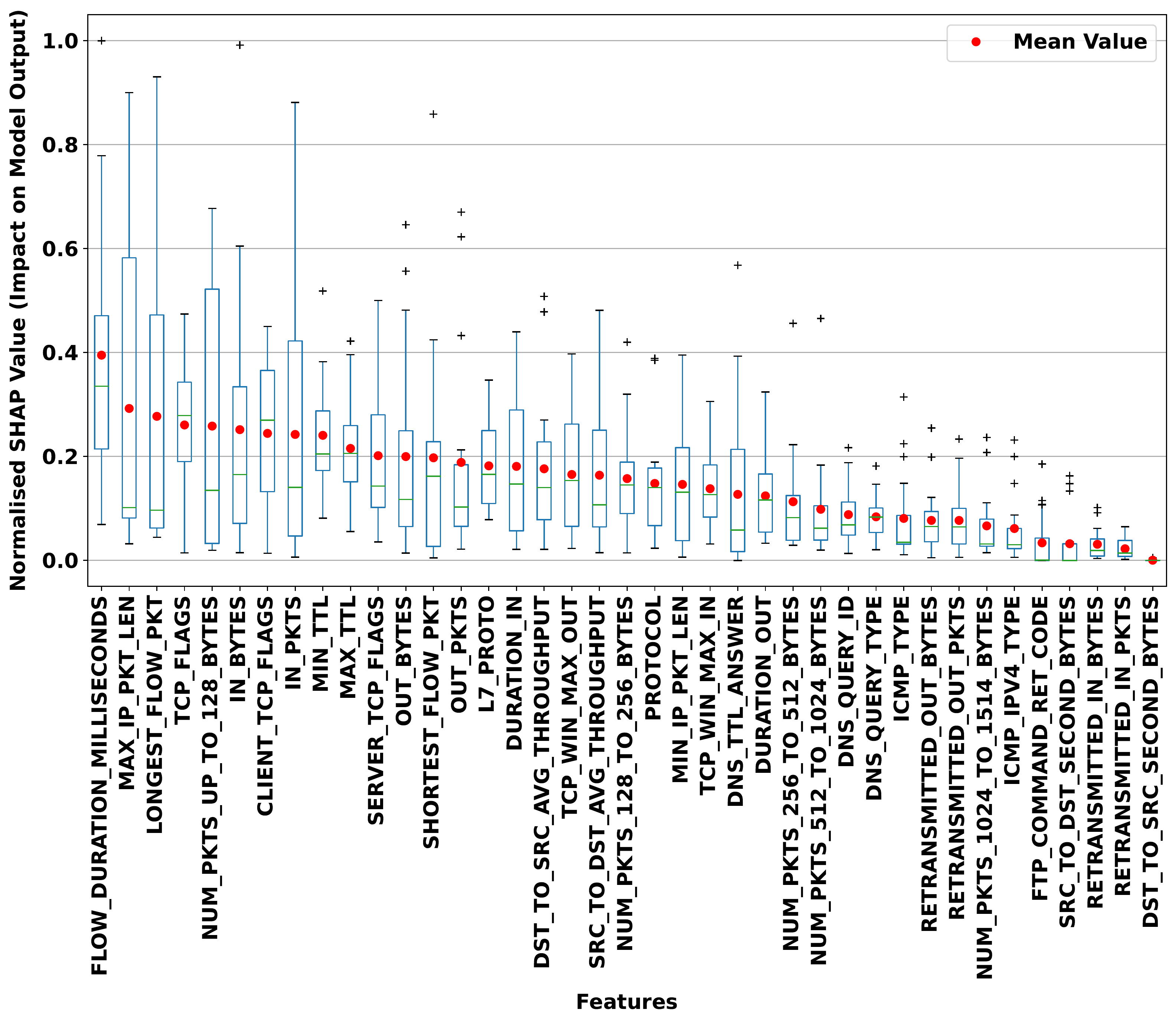}}%
    \vspace{\y}
    
    \qquad
    \subfloat[\centering ][oSVM]
    {\includegraphics[width=1\columnwidth, height=\x]{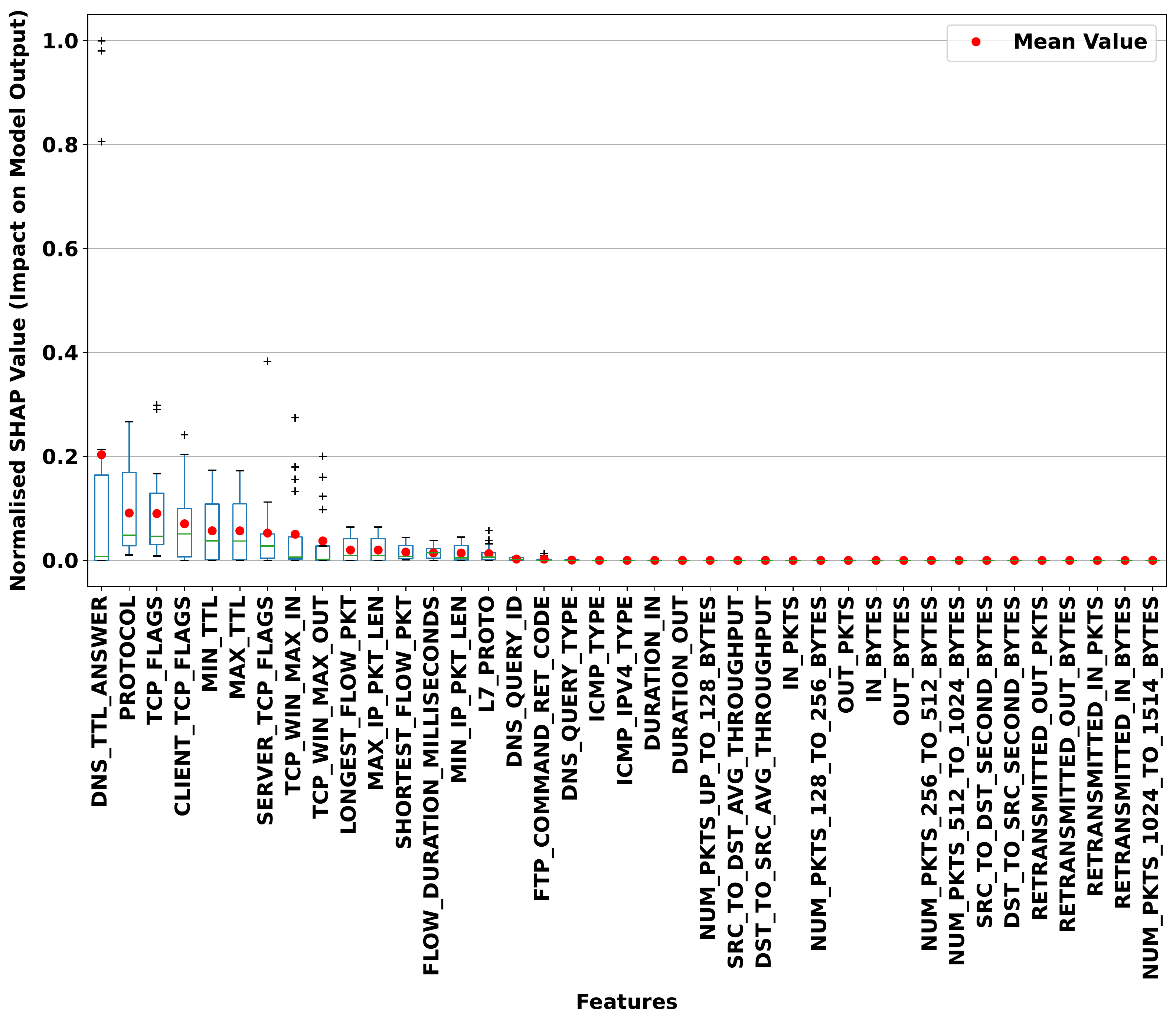}}%
    \vspace{\y}
    
    \qquad
    \subfloat[\centering ][SGD-oSVM]
    {\includegraphics[width=1\columnwidth, height=\x]{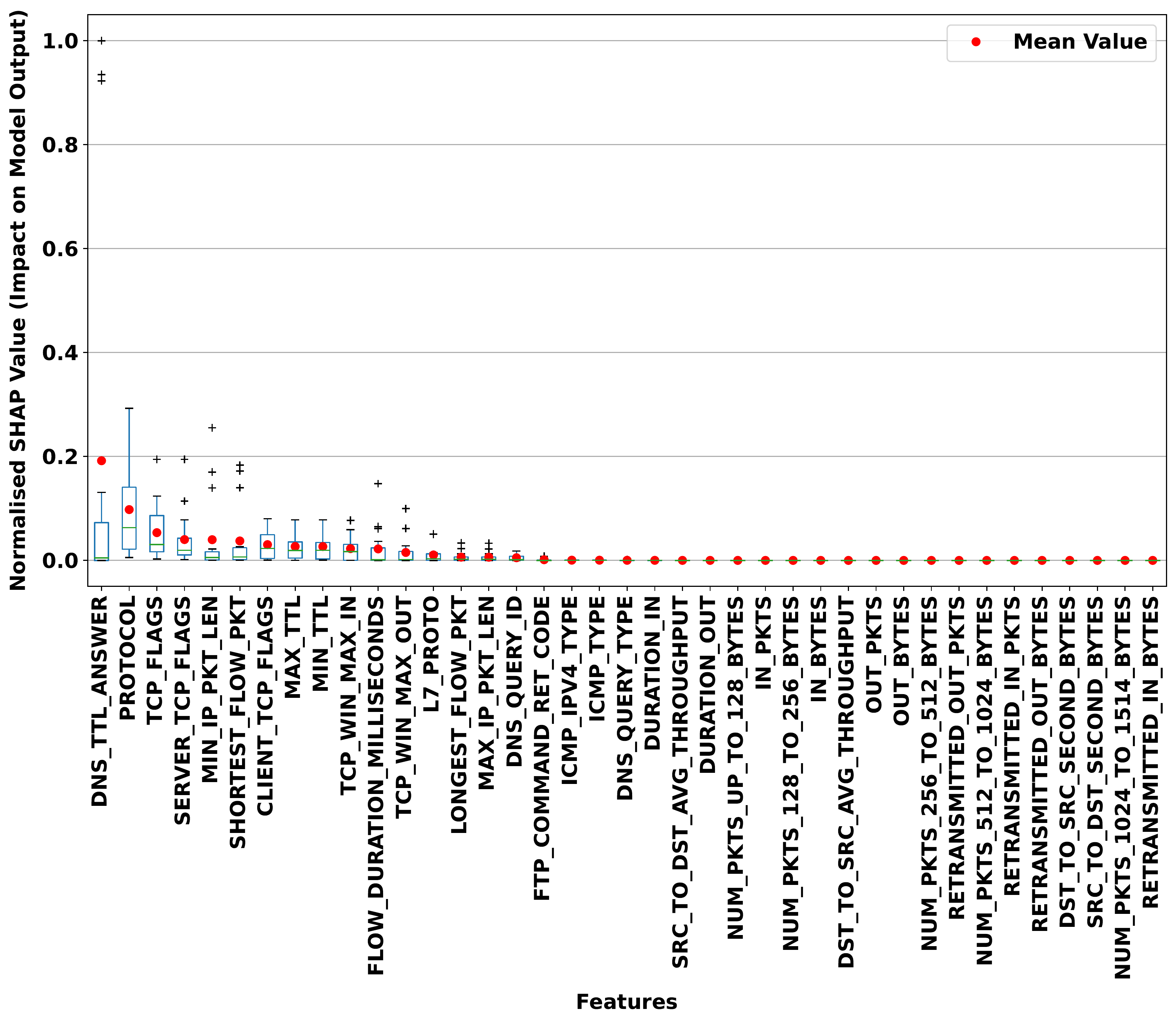}}%
    \vspace{\y}
    
    \caption{Distribution of mean SHAP values of features for (a) Isolation Forest, (b) oSVM and (c) SGD-oSVM models in 16 different experiments. Each Boxplot shows the range of mean SHAP values of a feature for the same model in 16 different combinations of the source-target domains.}%
    \label{fig:shap-boxbot}
\end{figure}
%

Accordingly, comparing the feature importance across all the 16 combinations of source-target domains will reveal the overall model behaviour. 
For this purpose, we computed the mean SHAP value of all features across all the experiments, and plotted their distributions side by side.
Figure~\ref{fig:shap-boxbot} shows the distribution of these mean SHAP values for the three unsupervised learning models, the Isolation Forest, oSVM and SGD-oSVM respectively plotted in  Figure~\ref{fig:shap-boxbot}-(a), (b), and (c).
The horizontal axis indicates the features and vertical axis indicates the normalized mean SHAP value.
Since the range of mean  SHAP values for different features were different, they have been normalized to make them comparable. 
The features are sorted in terms of their overall average in descending order, and the overall average of the mean SHAP value of each feature is also shown by a red circle. 

As can be seen, the variance of the mean SHAP values of features in the case of  Isolation Forest, Figure~\ref{fig:shap-boxbot}-(a), is significantly larger than the other two models, oSVM and SGD-oSVM shown in Figure~\ref{fig:shap-boxbot}-(b) and (c) respectively. 
As it was shown in Figure~\ref{fig:shap-mean}, the variations of feature order and importance values are directly linked to the variations of model behaviour.
Similarly, the considerable variations of the importance and order of the features across different source-target combinations for the Isolation Forest model indicates variations of its behaviour across these experiments.

The next two model, oSVM and SGD-oSVM, shown in Figure~\ref{fig:shap-boxbot}-(b) and (c) respectively, have much lower variations compared to Isolation Forest. 
As such, it is expected that their behaviour is more similar across different combinations of source-targets, which means these two model are more generalisable compared to Isolation  forest.
This can be easily verified by comparing the average decay per model as shown in Table~\ref{tab:unsupervised-decay}, which is $49.73\%$, $17.90\%$ and $17.36\%$ for the Isolation Forest, oSVM and SGD-oSVM, respectively.
The oSVM and SGD-oSVM models, in addition to the lower variance of mean SHAP values, share the same order/rank for many of the features.
This similarity in the distribution of mean SHAP value of features, which is an indicator of model behaviour similarity, explains their close average model decays ($17.90\%$ and $17.36\%$).


\begin{figure*}[!t]
\newcommand{\x}{4.25cm}
\newcommand{\z}{0.55cm}
    \centering
    \subfloat[\centering ][Feed Forward - source:UNSW, target:UNSW]
    {\includegraphics[width=1\columnwidth, height=\x]{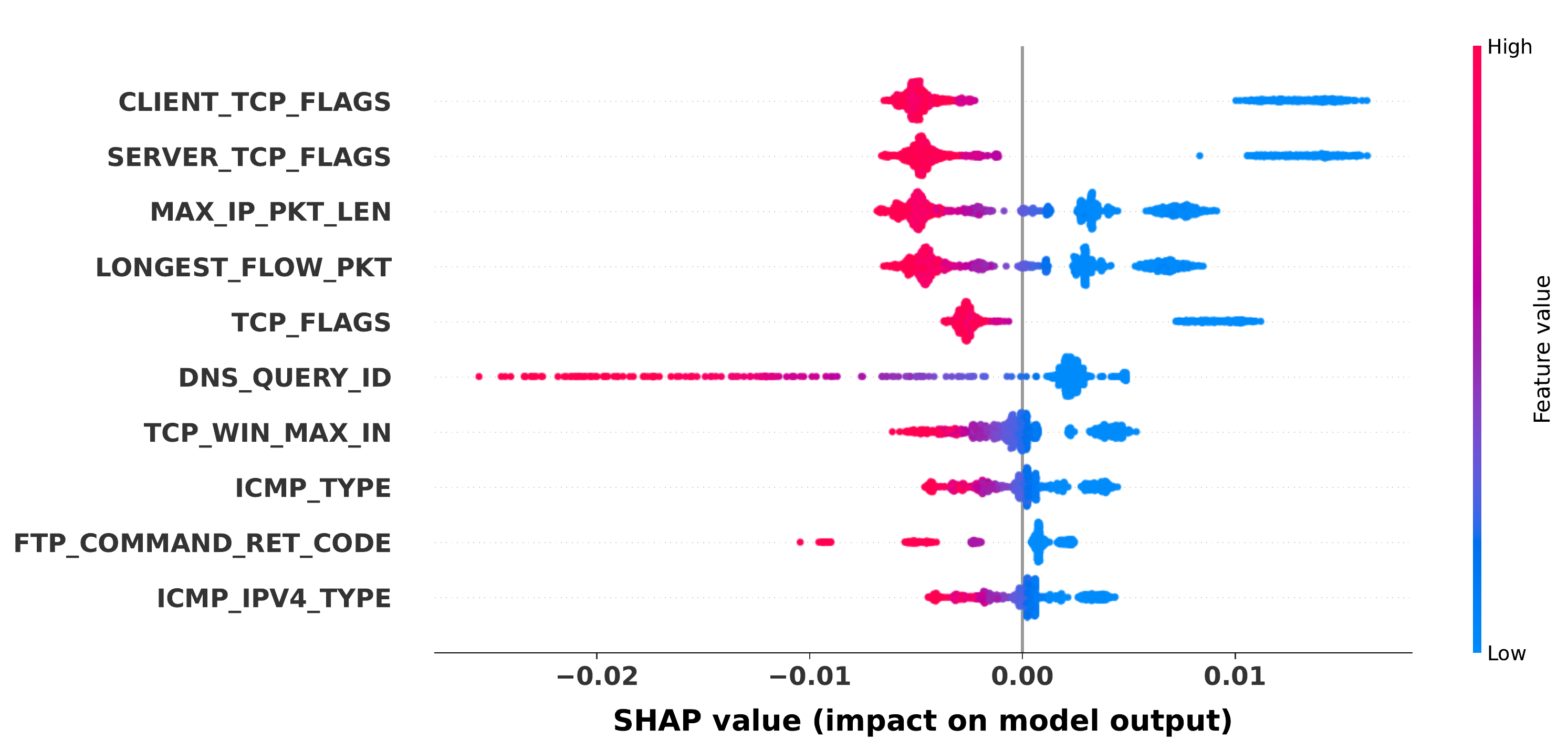}}%
    \hspace{0.5cm}
    \subfloat[\centering ][Feed Forward - source:UNSW, target:ToN]
    {\includegraphics[width=1\columnwidth, height=\x]{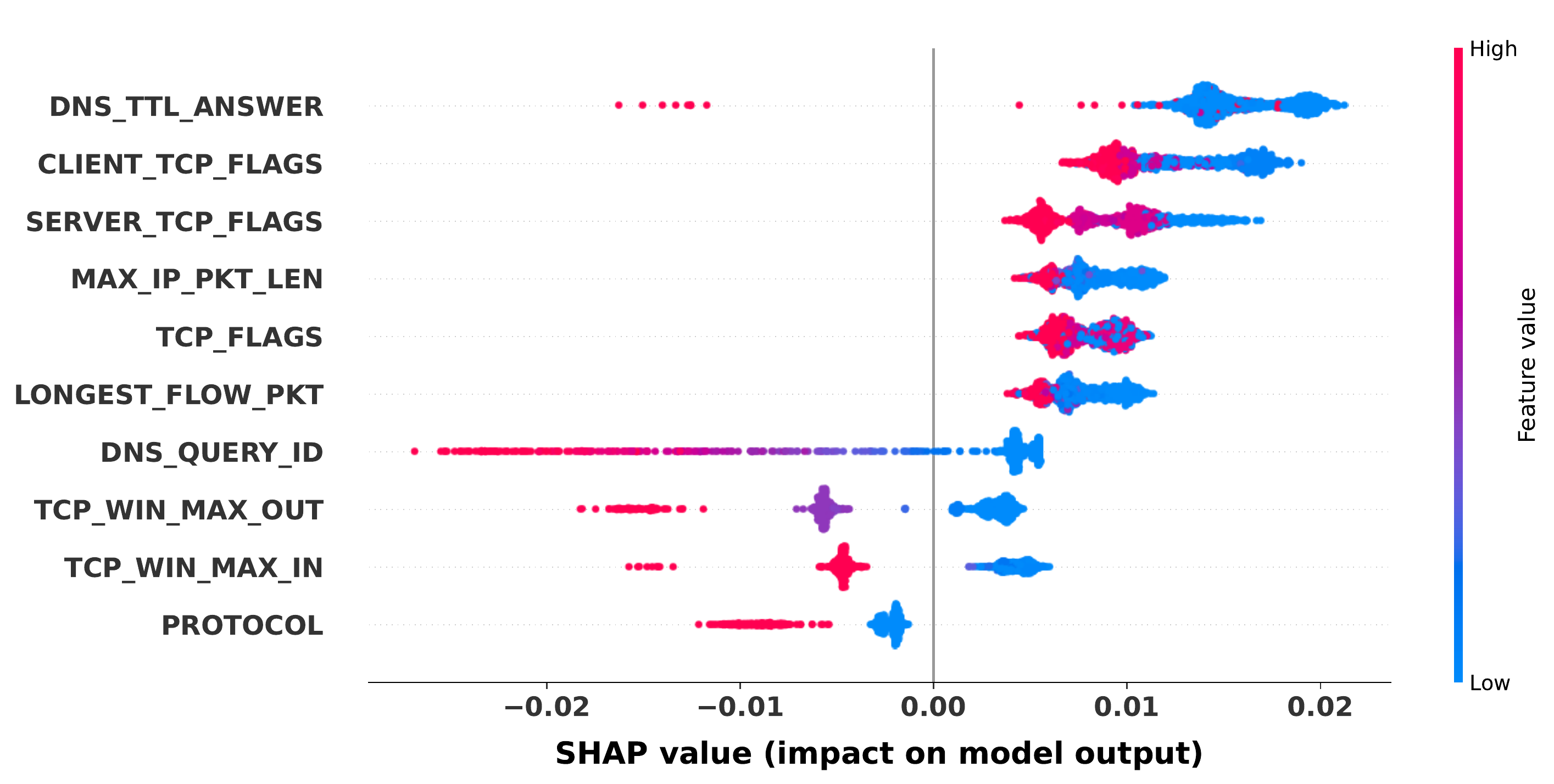}}%
    \vspace{0.5cm}
    \caption{SHAP summary plots for the top 10 features in the experiments  using the Feed Forward model where in (a) source: UNSW and target: UNSW  and in (b) source: UNSW and target: ToN.}%
    \label{fig:shap-summary-plot-1}
\end{figure*}

While the analysis of feature importance values explains the overall behaviour of the  models in terms of generalisability, it cannot answer a question like why a model performs well on one dataset and not well on another dataset. 
To answer this kind of questions about the behaviour of the models, we need a more detailed analysis of the SHAP values.

The SHAP summary plot seems an appropriate tool for this kind of analysis. 
A SHAP summary plot shows the SHAP value of features for individual data points. It shows how much impact each feature of a single data point has on generating the corresponding output. 
Figure~\ref{fig:shap-summary-plot-1}  shows two examples of SHAP summary plots.
In each SHAP summary plot the horizontal axis indicates the SHAP value, the vertical axis indicates the features, and the feature values are shown using the color range, from blue (low) to red (high).
A positive SHAP value in our experiments indicates the impact of the  feature towards the \textbf{Benign} class and a negative SHAP value indicates the impact towards the \textbf{Attack} class. 

Each point in this plot is created by two values, the feature and the SHAP values. 
Accordingly, a single instance of a dataset sample corresponds with the number of points equal to the number of features.
Since in these summary plots, only ten features are shown, a sample form a dataset corresponds with ten dots (data points) in a summary plot, one point per feature.
Wherever multiple samples have the same SHAP value, their representative dot-points are piled up in a histogram manner.
Hence, the larger height of the pile of dots indicates features of more  sample points have the same SHAP value.

Since it is not possible to investigate the SHAP summary plots of all the experiment pairs separately, we chose three pairs of experiments with the maximum contrasting results that explain the main aspects of generalisability observed in the results.  

The first aspect is the high performance of a model evaluated against the datasets it was trained on compared to the low performance of the same model when evaluated against other datasets (not seen during training).
Figure~\ref{fig:shap-summary-plot-1}-(a) and (b) are examples of SHAP summary plots for such a case in which the Feed Forward model is trained on NFv2-UNSW-NB15 dataset and is evaluated against (a) the NFv2-UNSW-NB15 dataset and (b) the NFv2-ToN-IoT dataset.

As can be seen, in Figure~\ref{fig:shap-summary-plot-1}-(a) the feature values have a clear correspondence with the Benign and Attack classes. Most of the Blue dots (low feature values) have positive SHAP values (Benign) and most of the Red dots (high feature values) have negative SHAP values (Attack). 
This is a simple multi-rule classifier that can separate the Attack and Benign classes based on the simple rules in terms of the feature values.

In Figure~\ref{fig:shap-summary-plot-1}-(b), however, this correspondence of the feature values with the Attack and Benign classes hardly can be seen and most of the feature values either indicate the Benign class or mixes of two classes. 
This is obviously due to the difference of the feature distribution in the new target domain, NFv2-ToN-IoT with the source domain NFv2-UNSW-NB15 dataset.
The result shown in Figure~\ref{fig:SE-trgt-summary}-(c) confirms our conclusion from the SHAP summary plots and show a significant performance drop for the this case.

\begin{figure*}[!b]
\newcommand{\x}{4.25cm}
\newcommand{\z}{0.55cm}
    \centering
    \subfloat[\centering ][Extra Tree - source:UNSW, target:BoT]
    {\includegraphics[width=1\columnwidth, height=\x]{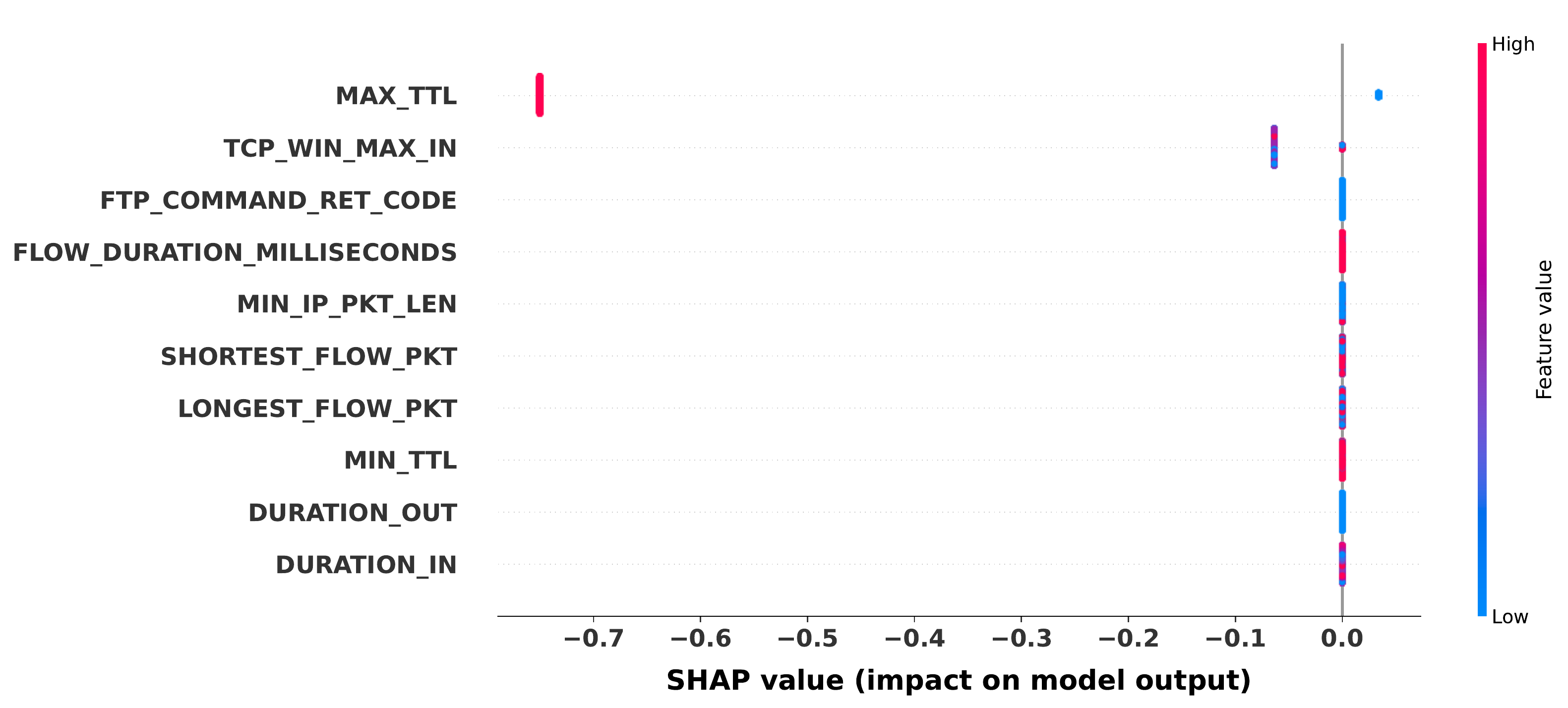}}%
    \hspace{0.5cm}
    \subfloat[\centering ][Extra Tree - source:BoT, target:UNSW]
    {\includegraphics[width=1\columnwidth, height=\x]{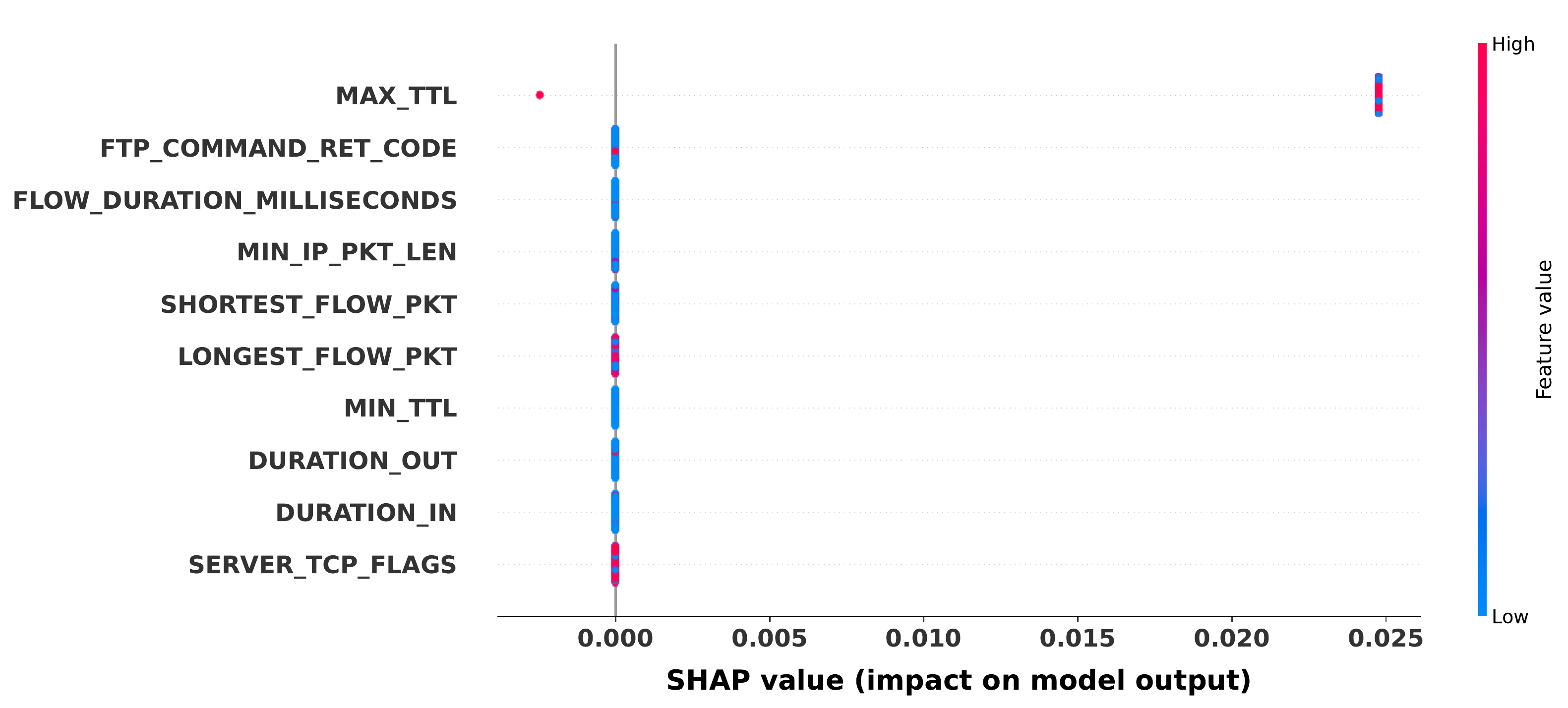}}%
   \vspace{0.5cm}
    \caption{SHAP summary plots for the top 10 features in the experiments using the Random Forest model where in (a) source: BoT and target: ToN, and in (b) source: ToN and target: BoT.}%
    \label{fig:shap-summary-plot-2}
\end{figure*}

The other aspect of generalisability observed in our results is its asymmetric behaviour, i.e., while a model has a high performance in a combination of the source-target domains, it has a low performance when the source and target domains are swapped.
Figure~\ref{fig:shap-summary-plot-2}-(a) and (b) illustrate an example of SHAP summary plots for such a case. The Extra Tree model in (a) is trained on NFv2-UNSW-NB15 dataset and evaluated on NFv2-BoT-IoT dataset, and in (b) it is trained on NFv2-BoT-IoT dataset and evaluated on NFv2-UNSW-NB15 dataset.

\begin{figure*}[!t]
\newcommand{\x}{4.25cm}
\newcommand{\z}{0.55cm}
    \centering
    \subfloat[\centering ][Extra Tree - source:BoT, target:ToN]
    {\includegraphics[width=1\columnwidth, height=\x]{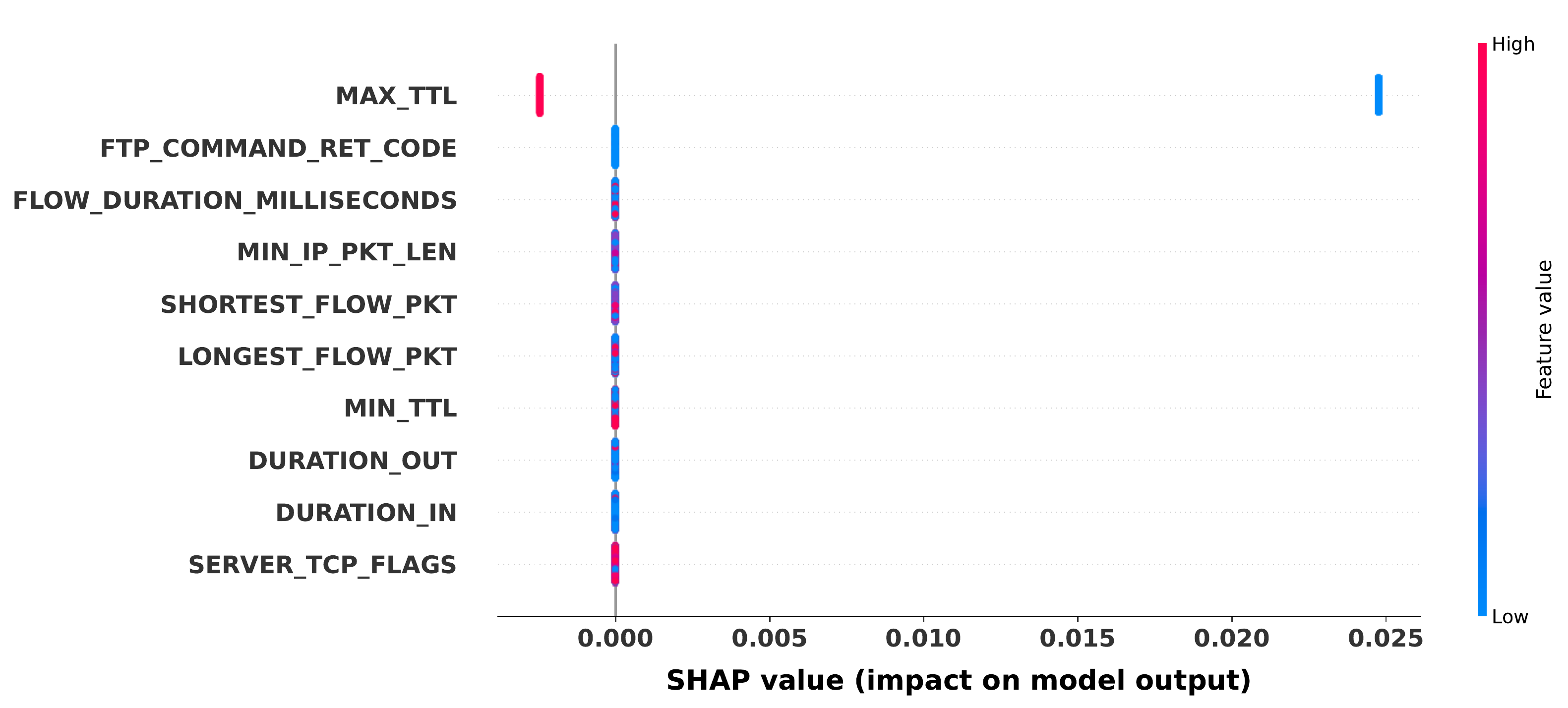}}%
    \hspace{0.5cm}
    \subfloat[\centering ][Extra Tree - source:ToN, target:BoT]
    {\includegraphics[width=1\columnwidth, height=\x]{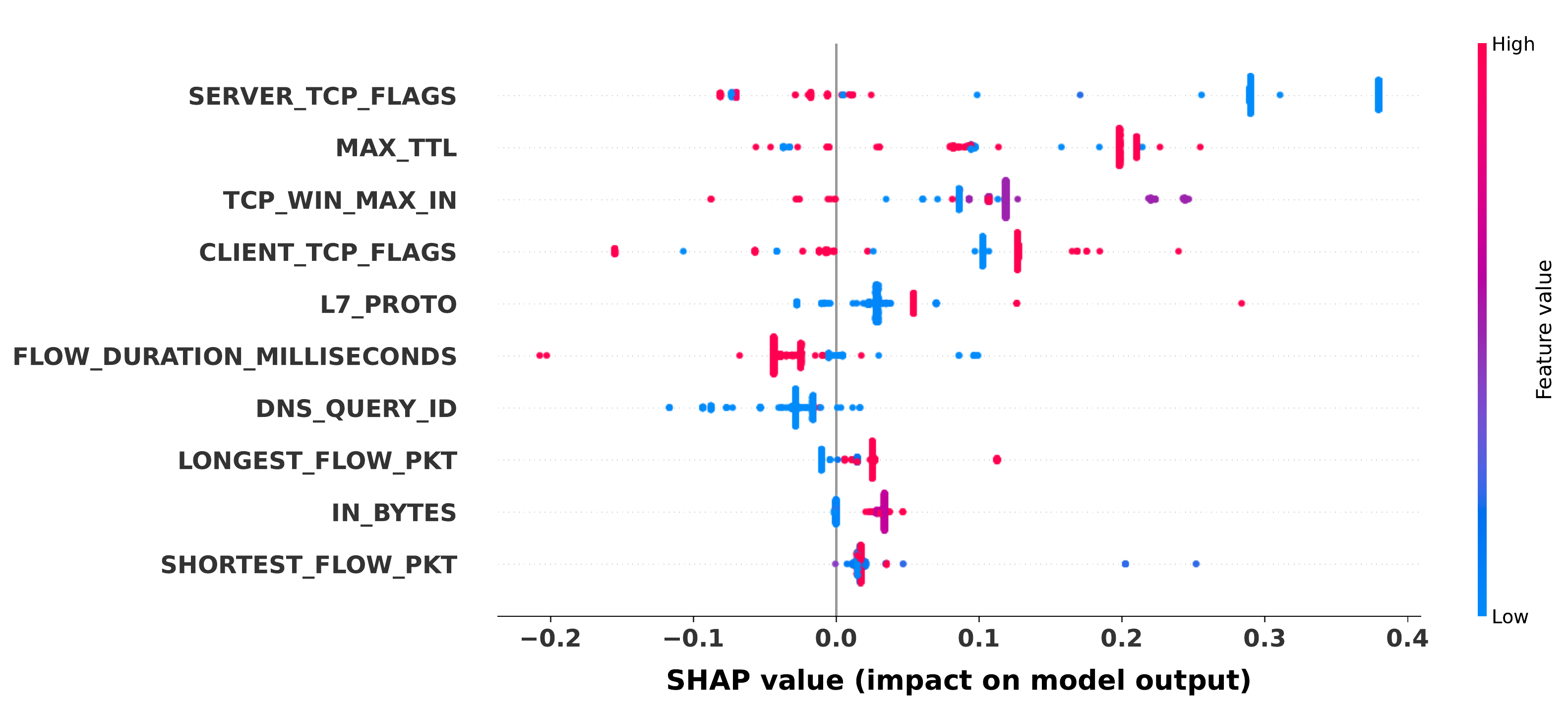}}%
    \vspace{\z}
    \caption{SHAP summary plots for the top 10 features in the experiments where in (a) source: BoT and target: ToN, and in (b) source: ToN and target: BoT, and in c) source: UNSW and target: BoT, and in (d) source: BoT and target: UNSW using Extra Tree model.}%
    \label{fig:shap-summary-plot-3}
\end{figure*}

As can be seen, in both SHAP summary plots most of the features have a zero SHAP value, indicating no impact on model output.
Though, in Figure~\ref{fig:shap-summary-plot-2}-(a) the low (blue) and high (red) values of the first important feature, MAX\_TTL, have a clear correspondence with the Benign and Attack classes (low feature values have positive and high feature values have negative SHAP values). 
However, in Figure~\ref{fig:shap-summary-plot-2}-(b) most of the MAX\_TTL values have positive SHAP values.
This means that the model has almost assigned all the MAX\_TTL values to the Benign class while other features have not been used in the classification at all. 

Hence, while a high performance is expected for the observed  \textit{single-rule classifier} in Figure~\ref{fig:shap-summary-plot-2}-(a) (with a huge difference between the SHAP values of the low and high MAX\_TTL values), it is hard to imagine a similar performance for the model in Figure~\ref{fig:shap-summary-plot-2}-(b) that has a single feature mostly indicating to the Benign class.
Both of these conclusions are consistent with the results shown in Figure~\ref{fig:SE-trgt-summary}-(a).

Finally, we investigate another case of asymmetric generalisability with a different distribution of SHAP values as shown in Figure~\ref{fig:shap-summary-plot-3}.
In Figure~\ref{fig:shap-summary-plot-3}-(a) the Extra Tree model is trained on NFv2-BoT-IoT dataset and evaluated on NFv2-ToN-IoT dataset, and in (b) it is trained on NFv2-ToN-IoT dataset and evaluated on NFv2-BoT-IoT dataset.

As can be seen in Figure~\ref{fig:shap-summary-plot-3}-(a) the SHAP values of all features except the MAX\_TTL is zero, i.e. they do not affect either of classes.
The dataset samples with a low MAX\_TTL value are classified as Benign and samples with a high MAX\_TTL value are classified as Attack.
In Figure~\ref{fig:shap-summary-plot-3}-(b), however, it is not possible to identify such a simple correspondence between the feature values and classes.
A mix of low and high feature values can be seen in the positive and negative ranges of the SHAP value.
Accordingly, it is expected that the model in Figure~\ref{fig:shap-summary-plot-3}-(a) that is equivalent to a simple single threshold classifier perform much better than the model in Figure~\ref{fig:shap-summary-plot-3}-(b). 
These conclusions are consistent with the results shown in Figure~\ref{fig:SE-trgt-summary}-(a).

\section{Conclusion}
Machine learning (ML) based network intrusion detection systems (NIDSs) have been around for many years to address the shortcomings of signature-based NIDSs.
While a large number of methods have been proposed in the academic literature of NIDS with near perfect detection and classification performances, the ML-based NIDSs rarely have been used in the real-world scenarios.
Since the evaluation of these methods is predominantly based on the datasets used for their training, we assume generalisability is the missing link to use these models in the real world applications. 

In this paper we extensively evaluate the generalisability of seven supervised and unsupervised ML-based NIDSs across four recently published publicly available NIDS datasets.
In these experiments, each ML-based NIDS is trained on a dataset and evaluated against all the four datasets, including the one used for its training.
This makes it possible to compare a model's performance in a single-domain and multi-domain (generalisability) evaluation.

The results indicate that while some models are able to generalise over one or two datasets, none of the studied models generalise well across all datasets.
The other observation is that generalisability can be asymmetric, which means performance of the model can significantly change when the source and target domains are swapped.
The last observation indicates that the unsupervised ML-based NIDSs generalise better than the supervised ML-based NIDSs, even though their single domain performance is lower than the supervised ML-based NIDSs.

We have further explained our results by finding the SHAP values for the model outputs.
Comparing the SHAP values of different dataset-model combinations indicates that the high classification performances in a combination of the model and source-target domains is mainly due to having one or more features that have strong correspondence with the Attack/Benign classes (positive and negative SHAP values).
Clearly, when such a simple rule (single or multi-threshold classifier) defines a model's performance, it can hardly be generalised to another target domain where features can have different distributions.
As such, the lack of generalisation of the model performance from one domain to the other, and the asymmetric behaviour of models generalisability can all be attributed to the presence of these simple rule classifiers for one combination of model-source-target datasets and its absence in other combinations due to different feature distributions.

As for our future research direction, we are investigating the ML-based approaches that can deal with/compensate for this feature distribution shifts in the alternate target domain, to enhance the generalisability of ML-based NIDSs.

\ifCLASSOPTIONcompsoc
  \section*{Acknowledgments}
\else
  \section*{Acknowledgment}
\fi
This research is made possible by an Advance Queensland Industry Research Fellowship, grant number RM2019002409.



%

\bibliography{main.bib}
\bibliographystyle{ieeetr}




\end{document}